\pdfoutput=1

\documentclass{sig-alternate}

\newcommand{\ignore}[1]{}
\usepackage{fancyhdr}
\usepackage[normalem]{ulem}
\usepackage[hyphens]{url}
\usepackage{hyperref}

\usepackage{amsmath}
\usepackage{mathtools}
\usepackage{verbatim}
\usepackage{color}
\usepackage{boxedminipage}
\usepackage{enumitem}
\usepackage[caption=false]{subfig}

\newcommand{\ie}{i.e.}
\newcommand{\etc}{etc.}
\newcommand{\eg}{e.g.}
\newcommand{\etal}{et al.}
\newcommand{\wrt}{w.r.t.}
\mathchardef\mhyphen="2D
\newcommand{\Hyphen}{\mhyphen}

\newcommand{\NmInst}{\mathsf{Nm}}
\newcommand{\LdInst}{\mathsf{Ld}}
\newcommand{\StInst}{\mathsf{St}}

\newcommand{\ComInst}{\mathsf{Commit}}
\newcommand{\RecInst}{\mathsf{Reconcile}}
\newcommand{\RMWInst}{\mathsf{RMW}}

\newcommand{\syncInst}{\mathsf{sync}}
\newcommand{\lwsyncInst}{\mathsf{lwsync}}
\newcommand{\isyncInst}{\mathsf{isync}}
\newcommand{\cmpInst}{\mathsf{cmp}}
\newcommand{\bcInst}{\mathsf{bc}}

\newcommand{\dmbInst}{\mathsf{dmb}}

\newcommand{\MBInst}{\mathsf{MEMBAR}}

\newcommand{\wmmDep}{WMM-D}
\newcommand{\oooNVP}{OOO-D}
\newcommand{\wmmSSB}{WMM-S}

\newcommand{\IIE}{$\mathrm{I^2E}$}

\newcommand{\True}{\mathsf{True}}
\newcommand{\False}{\mathsf{False}}
\newcommand{\ifFunc}{\mathbf{if}}
\newcommand{\thenFunc}{\mathbf{then}}
\newcommand{\elseFunc}{\mathbf{else}}
\newcommand{\whenFunc}{\mathsf{when}}
\newcommand{\decodeFunc}{\mathsf{decode}}
\newcommand{\executeFunc}{\mathsf{execute}}
\newcommand{\emptyFunc}{\mathsf{empty}}
\newcommand{\existFunc}{\mathsf{exist}}
\newcommand{\getYoungestFunc}{\mathsf{youngest}}
\newcommand{\enqFunc}{\mathsf{enq}}
\newcommand{\deqFunc}{\mathsf{deq}}
\newcommand{\getAnyAddr}{\mathsf{anyAddr}}
\newcommand{\insertFunc}{\mathsf{insert}}
\newcommand{\removeOldestFunc}{\mathsf{rmOldest}}
\newcommand{\removeOlderFunc}{\mathsf{rmOlder}}
\newcommand{\removeAddrFunc}{\mathsf{rmAddr}}
\newcommand{\getRandAndRemoveFunc}{\mathsf{getRandom}}
\newcommand{\clearFunc}{\mathsf{clear}}
\newcommand{\getRandFunc}{\mathsf{random}}
\newcommand{\getOldestFunc}{\mathsf{oldest}}
\newcommand{\hasTagFunc}{\mathsf{has}}
\newcommand{\maxFunc}{\mathsf{max}}
\newcommand{\letFunc}{\mathbf{let}}
\newcommand{\inFunc}{\mathbf{in}}
\newcommand{\decodeTSFunc}{\mathsf{decodeTS}}
\newcommand{\executeTSFunc}{\mathsf{executeTS}}

\newcommand{\noCycleFunc}{\mathsf{noCycle}}

\newcommand{\coOrd}{<_{co}}

\newcommand{\scNmRule}{SC-Nm}
\newcommand{\scLdRule}{SC-Ld}
\newcommand{\scStRule}{SC-St}

\newcommand{\tsoNmRule}{TSO-Nm}
\newcommand{\tsoLdRule}{TSO-Ld}
\newcommand{\tsoStRule}{TSO-St}
\newcommand{\tsoComRule}{TSO-Com}
\newcommand{\tsoDeqSbRule}{TSO-DeqSb}

\newcommand{\psoDeqSbRule}{PSO-DeqSb}

\newcommand{\wmmNmRule}{WMM-Nm}
\newcommand{\wmmLdSbRule}{WMM-LdSb}
\newcommand{\wmmLdMemRule}{WMM-LdMem}
\newcommand{\wmmLdIbRule}{WMM-LdIb}
\newcommand{\wmmStRule}{WMM-St}
\newcommand{\wmmRecRule}{WMM-Rec}
\newcommand{\wmmComRule}{WMM-Com}
\newcommand{\wmmDeqSbRule}{WMM-DeqSb}

\newcommand{\wmmDepNmRule}{\wmmDep-Nm}
\newcommand{\wmmDepStRule}{\wmmDep-St}
\newcommand{\wmmDepLdSbRule}{\wmmDep-LdSb}
\newcommand{\wmmDepLdIbRule}{\wmmDep-LdIb}
\newcommand{\wmmDepLdMemRule}{\wmmDep-LdMem}
\newcommand{\wmmDepRecRule}{\wmmDep-Rec}
\newcommand{\wmmDepComRule}{\wmmDep-Com}
\newcommand{\wmmDepDeqSbRule}{\wmmDep-DeqSb}

\newcommand{\wmmSSBPropSt}{\wmmSSB-Copy}
\newcommand{\wmmSSBStRule}{\wmmSSB-St}
\newcommand{\wmmSSBDeqSbRule}{\wmmSSB-DeqSb}

\newcommand{\rcLdBeforeStCons}{RC-LdVal-1}
\newcommand{\rcStBeforeLdCons}{RC-LdVal-2}

\newcommand{\rcDepCons}{RC-Dependency}
\newcommand{\rcCoCons}{RC-Coherence}
\newcommand{\rcNoDLCons}{RC-No-Deadlock}
\newcommand{\rcFixLdValCons}{RC-Fix-WC-Ld-Value}
\newcommand{\rcFixDepCons}{RC-Fix-Dependency}

\newcommand{\rmoFixLd}{RMO-Fix-Ld}

\newcommand{\reduceRuleSpace}{\vspace{-5pt}}
\newcommand{\reduceRuleEndSpace}{\vspace{-5pt}}

\setlist{noitemsep, leftmargin=*,topsep=3pt}

\fancypagestyle{firstpage}{
	\fancyhf{}
	
	\pagenumbering{arabic}
}

\def\sharedaffiliation{%
\end{tabular}
\begin{tabular}{c}}

\title{Taming Weak Memory Models\vspace{-30pt}} 

\author{
	Sizhuo Zhang \\ 
	\email{szzhang@mit.edu} 
	\and
	Arvind \\
	\email{arvind@csail.mit.edu}
	\and
	Muralidaran Vijayaraghavan \\
	\email{vmurali@csail.mit.edu}
	\sharedaffiliation
	{Department of Electrical Engineering and Computer Science} \\
	{Massachusetts Institute of Technology}
}

\begin{document}
\maketitle
\thispagestyle{firstpage}
\pagestyle{plain}

\begin{abstract}

Speculative techniques in microarchitectures relax various dependencies in programs, which contributes to the complexity of (weak) memory models. We show using WMM, a new weak memory model, that the model becomes simpler if it includes load-value speculation and thus, does not enforce any dependency! 
However, in the absence of good value-prediction techniques, a programmer may end up paying a price for the extra fences. 
Thus, we also present \wmmDep, which enforces the dependencies captured by the current microarchitectures. 
\wmmDep{} is still much simpler than other existing models. 
We also show that non-atomic multi-copy stores\footnote{Also known as non-multi-copy-atomic stores \cite{lustig2015armor}.} arise as a result of sharing write-through caches. 
We think restricting microarchitectures to write-back caches (and thus simpler weak memory models) will not incur any performance penalty.
Nevertheless, we present \wmmSSB, another extension to WMM, which could model the effects of non-atomic multi-copy stores.

WMM, \wmmDep{}, and \wmmSSB{} are all defined using \emph{Instantaneous Instruction Execution} (\IIE), a new way of describing memory models without explicit reordering or speculative execution.

\end{abstract}

\section{Introduction} \label{sec: intro}

Architects have often made changes for performance in microarchitectures, which end up affecting the memory model of multiprocessors in subtle ways. 
The problem is serious because the official definitions of memory models, such as Power and ARM, are underspecified in the company documents \cite{power2013version,armv7ar}.
In order to resolve the ambiguity in the official documents, researchers have used additional empirical evidence and developed abstract machines and axioms to capture the behaviors of these models precisely. 
This situation is not satisfactory either, because these formal definitions use complex abstract machines and specify a large number of axioms to capture all the corner cases \cite{sarkar2011understanding,mador2012axiomatic,flur2016modelling}.
Additionally, most architects are intimidated by these formalizations.

If the behaviors to be modeled are complicated, then there is no reason to believe that the model definition can be simple.
Indeed we believe this is the case with Power and ARM models, because they must capture the non-atomic multi-copy stores.
Since these types of stores arise as a consequence of sharing write-through caches, one way to simplify the behaviors is to use write-back caches exclusively.
Another source of complexity comes from the fact that speculative execution selectively relaxes program dependencies.
It is complicated to model precisely the forbidden behaviors that are caused by specific dependencies.

Sometimes memory models have been defined \emph{intuitively but incorrectly}.
We will show that the RC definition \cite{gharachorloo1990memory} is imprecise both in dependency ordering and the values returned by loads, and that it is not easy to fix the model without making it too restrictive.
We will also show that the RMO definition \cite{weaver1994sparc} is fairly precise, but rules out certain architectural optimizations which are commonly believed to be permitted.

We want a memory model that simultaneously satisfies the following three criteria:
\begin{enumerate}
	\item Model definition is simple and precise;
	\item Model allows efficient hardware implementations;
	\item High-level language (HLL) primitives can be efficiently mapped to memory instructions and fences in the model.
\end{enumerate}
By \emph{simple} definition, we mean an \emph{Instantaneous Instruction Execution} (\IIE{}) description of the instruction set.
All abstract machines in the \IIE{} framework consist of a monolithic multi-ported memory and $n$ atomic processors. 
The atomic processor executes each instruction instantaneously so the architecture state is by definition always up-to-date.
The instruction reordering and dependencies in specific models can be captured by including different types of buffers between each processor and its port in the memory.
\IIE{} descriptions are considerably simpler than other model definitions which either use reordering axioms or operational models that execute instructions partially.

The third criterion is important because people prefer to implement concurrent programs using primitives in HLLs instead of architecture-specific fences in assembly, and this requires a systematic mapping from HLL semantics to weak memory models.

None of the current architectural memory models satisfy all the three criteria.
SC and TSO have simple and precise descriptions, but leave a lot on the table as far as performance is concerned.
As already mentioned, weak memory models like Power, ARM, RC, and RMO do not satisfy the first criterion.
The situation is further complicated by the fact that there are attempts to design HLLs and extend weak models so that the two would be more compatible.

The major contributions of this paper are:
\begin{enumerate}
	\item Identification of architectural features that cause non-atomic multi-copy stores in Power and ARM models;
	\item The first characterization of the problems with RC and RMO definitions;
	\item \IIE, a new framework for describing memory models;
	\item WMM, an \IIE{} weak memory model which satisfies the three criteria of goodness given earlier, and its implementation with speculations on all dependencies;
	\item \wmmDep{}, an extension of WMM to capture data-dependency ordering precisely while still using \IIE{}.
	\item \wmmSSB, an extension of WMM to model non-atomic multi-copy stores.
\end{enumerate}

\noindent\textbf{Paper organization:}
Section \ref{sec: related work} presents the related work.
We analyze the source of complexity in Power and ARM memory models in Section \ref{sec: power arm}.
We identify the problems in the definitions of RC and RMO in Section \ref{sec: rc rmo}.
In Section \ref{sec: I2E}, we propose a new framework called \textbf{\IIE} to define memory models.
We use \IIE{} to give simple abstract machines of SC, TSO and PSO.
We define WMM using \IIE{} in Section \ref{sec: WMM}, which also includes its implementation, litmus tests and a compilation scheme from C++.
We define \wmmDep{} to capture data dependency in Section \ref{sec: data dep}.
In Section \ref{sec: non atomic mem}, we extend WMM to \wmmSSB{} to model non-atomic multi-copy stores. 
Our conclusions are presented in Section \ref{sec: conclude}.

\section{Related Work} \label{sec: related work}

SC \cite{lamport1979make} is the most intuitive memory model, but naive implementations of SC suffer from poor performance.
Gharachorloo \etal{} proposed load speculation and store prefetch to enhance the performance of SC \cite{gharachorloo1991two}.
Over the years, researchers have proposed more aggressive techniques to preserve SC \cite{ranganathan1997using,guiady1999sc+,gniady2002speculative,ceze2007bulksc,wenisch2007mechanisms,blundell2009invisifence,singh2012end,lin2012efficient,gope2014atomic}. 
Perhaps because of their hardware complexity, the adoption of these techniques in commercial microprocessor has been limited. 
Instead the manufactures and researchers have chosen to present weaker memory model interfaces, \eg{} TSO \cite{sparc1992sparcv8}, PSO \cite{weaver1994sparc}, RMO \cite{weaver1994sparc}, Processor Consistency \cite{goodman1991cache}, Weak Consistency (WC) \cite{dubois1986memory}, RC \cite{gharachorloo1990memory}, CRF \cite{shen1999commit}, Power \cite{power2013version} and ARM \cite{armv7ar}.
The tutorials by Adve \etal{} \cite{adve1996shared} and by Maranget \etal{} \cite{maranget2012tutorial} provide relationships among some of these models.

The lack of clarity in the definitions of Power and ARM memory models in their respective company documents has led some researchers to empirically determine allowed/disallowed behaviors \cite{sarkar2011understanding,mador2012axiomatic,alglave2014herding,flur2016modelling}.
Based on such observations, in the last several years, both \emph{axiomatic} models and \emph{operational} models have been developed which are compatible with each other \cite{alglave2009semantics,Alglave2011,alglave2012formal,mador2012axiomatic,alglave2014herding,sarkar2011understanding,sarkar2012synchronising,alglave2013software,flur2016modelling}.
However, these models are quite complicated; for example, the Power axiomatic model has 10 relations, 4 types of events per instruction, and 13 complex axioms \cite{mador2012axiomatic}, some of which have been added over time to explain specific behaviors \cite{alglave2009semantics,Alglave2011,alglave2012fences,mador2012axiomatic}. 
The abstract machines used to describe Power and ARM operationally are also quite complicated, because they require the user to think in terms of partially executed instructions~\cite{sarkar2011understanding,sarkar2012synchronising}.
In particular, the processor sub-model incorporates ROB operations, speculations, instruction replay on speculation failures, \etc{}, explicitly, which are needed to explain the enforcement of specific dependency (\ie{} data dependency).
We present an \IIE{} model \wmmDep{} in Section \ref{sec: data dep} that captures data dependency and sidesteps all these complications.
Another source of complexity is in the memory sub-model, which we explain in Section \ref{sec: power arm}.

Adve \etal{} defined Data-Race-Free-0 (DRF0), a class of programs where shared variables are protected by locks, and proposed that DRF0 programs should behave as SC \cite{adve1990weak}.
However, architectural memory models must also define the behaviors of non-DRF0 programs.

A large amount of research has also been devoted to specifying the memory models of HLLs: C++ \cite{c++n4527,boehm2008foundations,batty2011mathematizing}, Java \cite{manson2005java,cenciarelli2007java, maessen2000improving}, \etc{}
We will provide compilation schemes from C++, a widely-used HLL, to the WMM and \wmmDep{} models presented in this paper.

Arvind and Maessen have specified precise conditions for preserving store atomicity in program execution even when instructions can be reordered  \cite{arvind2006memory}. 
In contrast, the WMM and \wmmDep{} models presented in this paper do not insist on store atomicity at the program level. 

Recently, Lustig \etal{} have used Memory Ordering Specification Tables (MOSTs) to describe memory models, and proposed a hardware scheme which dynamically converts programs across memory models described in MOSTs \cite{lustig2015armor}.
MOST specifies the ordering strength (\eg{} locally ordered, multi-copy atomic) of two instructions from the same thread under different conditions (\eg{} data dependency, control dependency).
It is not clear to us what events in the program execution are being (re)ordered by MOST. 
It is also unclear regarding which value a load returns given a legal order of events.

\section{Non-atomic multi-copy stores: an Avoidable Complication} \label{sec: power arm}

Stores in TSO (Intel) are known as \emph{multi-copy atomic}, because a store first becomes visible to the local processor and then later to all other processors simultaneously.
In contrast, Stores in Power and ARM processors are \emph{non-atomic multi-copy}, that is, a store may become visible to different processors at different times.
This is caused by the memory system, which allows a store $S$ from processor $Pi$ to be observed by $Pj$ before $S$ has finished all coherence transactions. 
There are two root causes for this behavior.
If multiple threads on a single core, \ie{} as in Simultaneous Multithreading (SMT), share a store buffer, then a store by any of these threads may be seen by all these threads before other processors.
This non-atomicity can be avoided by keeping the stores of threads separate by tagging them with thread IDs in the store buffer.
If such tagged shared store buffers are combined with write-back caches, all threads except the one which issued the store cannot observe this store until the store is committed to L1.
On the other hand, if multiple threads share a write-through cache (typically L1), then these threads can see a store by any of these threads before the store reaches coherence, \ie{} become globally visible, making it non-atomic.
Unlike the case of shared store buffer, it is infeasible to distinguish between stores by different threads in the write-through cache.
Even without SMT, the non-atomicity problem will persist if the shared L2 is write-through.

Sorin \etal{} have identified that it is hard to implement TSO with shared write-through caches \cite[page 180]{sorin2011primer}.
The above analysis also matches their understanding.
Later we will show that a weak memory model with multi-copy atomic stores, is considerably simpler than Power and ARM models. 
In case one really wants to model a non-atomic multi-copy memory system, we also present such a model using \IIE{} in Section \ref{sec: non atomic mem}.

\section{Incorrect Memory Model Definitions} \label{sec: rc rmo}

It is generally believed that RC and RMO are well-defined memory models.
We first show that the RC definition \cite{gharachorloo1990memory}, in fact, is underspecified in the sense that it does not precisely define the values returned by loads, and the ordering of events in case of dependent instructions.
We will further show that after several attempts to resolve the ambiguity, the resulting model is no longer weak enough.
RMO definition \cite{weaver1994sparc}, on the other hand, has a precise axiomatic description, but it fails to match an architect's intuitive understanding of the model.
(Readers could skip this section without losing continuity, though later sections will refer to program examples used here).

\subsection{RC}
\subsubsection{Original RC Definition \cite{gharachorloo1990memory}}
A memory access in RC has an ``\emph{issue}" event followed by $n$ ``\emph{being performed with respect to (\wrt{}) processor $i$}" ($i=1\ldots n$) events.
The order of ``issue" and ``being performed \wrt{}" events constrain the load values in two ways (see Definition 2.1 in \cite{gharachorloo1990memory}):
\begin{itemize}
	\item \emph{\rcLdBeforeStCons}: If load $L$ is performed \wrt{} $Pi$ before store $S$ is issued by $Pi$, $L$ cannot read from $S$.
	\item \emph{\rcStBeforeLdCons}: If store $S$ to address $a$ is performed \wrt{} $Pi$ before load $L$ to $a$ is issued by $Pi$, $L$ must read from $S$ or another store $S'$ which is performed \wrt{} $Pi$ after $S$.
\end{itemize}
When a memory access is performed \wrt{} all processors, the access is \emph{performed}.
Notice that Definition 2.1 of \cite{gharachorloo1990memory} is ambiguous about the load result in case that $L$ is issued before $S$ is performed while $L$ is performed after $S$ is performed.
Thus, our attempted interpretation, \rcStBeforeLdCons, only constrains the load result in case $L$ is issued after $S$ is performed.

RC classifies memory accesses into two categories: \emph{ordinary} ones and \emph{special} ones.
Special accesses are further partitioned into three types: \emph{acquire} ($\mathsf{acq}$), \emph{release} ($\mathsf{rel}$) and \emph{non-synchronization}.
RC requires ordinary accesses to be \emph{performed} before or after release and acquire accesses, respectively (see Condition 3.1 in \cite{gharachorloo1990memory}). 
In addition, RC places the following three constraints (see the last paragraph of Section 2 of \cite{gharachorloo1990memory}):
\begin{itemize}
	\item \emph{\rcDepCons}: ``Uniprocessor control and data dependences are respected".
	\item \emph{\rcCoCons}: ``All writes to the same location are serialized in some order and are performed in that order \wrt{} any processor".
	The last store in that order gives the final memory value for that location.
	\item \emph{\rcNoDLCons}: ``Accesses that occur previously in program order eventually get performed".
\end{itemize}
A legal execution is a total order of all events which satisfies all the constraints.

\subsubsection{Ambiguity in RC Definition}
We explain the ambiguity using examples.
All examples in this paper assume that all memory locations are initialized to 0.

\noindent\textbf{Load results}:
Consider the program in Figure \ref{fig: rc mp}, which is a common usage of acquire-release pair to communicate data.
The behavior in the figure is allowed by the above definition, which is unexpected. 
Consider the following event order: $I_1$ to $I_4$ are issued one by one, and then $I_1$ to $I_4$ are performed one by one.
Since there is no dependency in the program, events can be issued at the beginning, and the above order is legal.
Since both $I_1$ and $I_4$ are issued before any of them is performed \wrt{} any processor, neither \rcLdBeforeStCons{} nor \rcStBeforeLdCons{} could constrain $I_4$ to only read from $I_1$.
Thus $I_4$ is allowed to read the initial value 0.

\begin{figure}[!htb]
	\begin{minipage}{0.49\columnwidth}
		\small
		\begin{tabular}{|l|l|}
			\hline
			Proc. P1 & Proc. P2 \\
			\hline
			$\!\! I_1: \StInst\ a\ 1 \!\!$       & $\!\! I_3: r_1=\LdInst_{acq}\ b \!\!$  \\
			$\!\! I_2: \StInst_{rel}\ b\ 1 \!\!$ & $\!\! I_4: r_2=\LdInst\ a \!\!$ \\
			\hline
			\multicolumn{2}{|l|}{RC allows: $r_1=1,\ r_2=0$} \\
			\hline
		\end{tabular}
		\caption{Message passing in RC} \label{fig: rc mp}
	\end{minipage}
	\hspace{1pt}
	\begin{minipage}{0.4\columnwidth}
		\small
		\begin{tabular}{|l|l|}
			\hline
			Proc. P1 & Proc. P2 \\
			\hline
			$\!\! I_1: r_1=\LdInst\ a \!\!$  & $\!\! I_3: r_2=\LdInst\ b \!\!$ \\
			$\!\! I_2: \StInst\ b\ r_1 \!\!$ & $\!\! I_4: \StInst\ a\ r_2 \!\!$ \\
			\hline
			\multicolumn{2}{|l|}{RC allows: $r_1=r_2=42$} \\
			\hline
		\end{tabular}
		\caption{Thin-air read in RC} \label{fig: rc thin air}
	\end{minipage}
\end{figure}

\noindent\textbf{Ambiguous dependency constraint}:
There are two interpretations of the \rcDepCons{} constraint. Suppose it means that if an access $B$ depends on another access $A$, then $B$ cannot be issued before $A$ is performed. 
Obviously with this interpretation, no speculative execution is possible.

Another interpretation is that dependency does not enforce any ordering of events, as long as the final result satisfies the program logic.
This interpretation is so relaxed that it allows the ``thin-air read" behavior as shown in Figure \ref{fig: rc thin air}.
This behavior may incur security problems, and is explicitly forbidden by C++ \cite{c++n4527} for relaxed atomic loads and stores.
Thus the compilation from C++ to RC will be inefficient.

RC definition is also unclear about how two loads to the same address on the same processor affect the ordering of events.

\subsubsection{Attempts to Fix RC}

Since RC is an extension for WC, suppose we constrain the load values by borrowing Dubois' \etal{} \cite{dubois1986memory} WC solution: 
\begin{itemize}
	\item \emph{\rcFixLdValCons}: The result of a load on $Pi$ should be the value given by the latest store (for the same address) performed \wrt{} $Pi$.
\end{itemize}
With \rcFixLdValCons{}, when a load gets its result by reading from a store, the load is automatically performed \wrt{} all processors.
(It is easy to see that \rcFixLdValCons{} implies \rcLdBeforeStCons{} and \rcStBeforeLdCons{}, so we do not need these two earlier conditions). 

We can fix the dependency constraint to allow speculation while avoiding ``thin-air" read as follows:
\begin{itemize}
	\item\emph{\rcFixDepCons}: A store $S$ that depends on a load $L$ should not be performed \wrt{} any other processor before $L$ is performed.
\end{itemize}

Now consider the Write-Write Causality (WWC) program in Figure \ref{fig: rc no ssb}. 
According to the RC definition after fixes, $I_1$ will be performed before $I_5$ \wrt{} P2, thus making it impossible for the $m[a]$ to be 2 according to \rcCoCons{}. 
However, most architects believe that RC can be implemented with a non-atomic multi-copy memory system (\eg{} P1 and P2 shares a write-through L1), which allows this behavior.

Furthermore, if we move $I_1$ from P1 to P2 and place $I_1$ right before $I_2$ in Figure \ref{fig: rc no ssb}, the behavior in the figure is still disallowed by RC for the same reason.
However, the behavior is observable in implementations, in which P2 could locally forward the data of $I_1$ to $I_2$. 
And such implementations are also believed to be permitted by RC.
Thus, the gap between the intuitive understanding of RC and its precise definition remains. 
It is probably possible to come up with a much more complex fix for RC, but then it will not be a simple model anymore.

\begin{figure}[!htb]
	\centering
	\small
	\begin{tabular}{|l|l|l|}
		\hline
		Proc. P1 & Proc. P2 & Proc. P3 \\
		\hline
		$I_1: \StInst\ a\ 2$ & $I_2: r_1=\LdInst\ a$        & $I_4: r_2=\LdInst\ b$     \\
		                     & $I_3: \StInst\ b\ (r_1 - 1)$ & $I_5: \StInst\ a\ r_2$     \\
		\hline
		\multicolumn{3}{|l|}{RC forbids: $r_1=2,\ r_2=1,\ m[a]=2$} \\
		\hline
	\end{tabular}
	\caption{WWC (non-atomic multi-copy stores) in RC} \label{fig: rc no ssb}
\end{figure}

\subsection{RMO} \label{sec: rmo}
There is also a gap between the definition of RMO \cite[Appendix D]{weaver1994sparc} and the optimizations which are expected to be allowed in its implementation.
RMO forbids the behavior shown in Figure \ref{fig: rmo spec} because $I_9$ transitively depends on $I_4$ according to the definition, and $I_9$ is ordered after $I_4$ in memory order.
However, this behavior is possible in hardware with speculative load execution and store forwarding, \ie{} $I_7$ first speculatively bypasses from $I_6$, and then $I_9$ speculatively executes to get 0.
Most architects will not be willing to give up on these two optimizations.

\begin{figure}[!htb]
	\begin{minipage}[b]{0.55\columnwidth}
		\centering
		\small
		\begin{tabular}{|l|l|}
			\hline
			Proc. P1 & Proc P2 \\
			\hline
			$\!\! I_1: \StInst\ a\ 1 \!\!$ & $\!\! I_4: r_1=\LdInst\ b \!\!$ \\
			$\!\! I_2: \MBInst \!\!$       & $\!\! I_5: \ifFunc(r_1 \!\neq\! 1)\ \mathsf{exit} \!\!$ \\
			$\!\! I_3: \StInst\ b\ 1 \!\!$ & $\!\! I_6: \StInst\ c\ 1 \!\!$ \\
			                               & $\!\! I_7: r_2=\LdInst\ c \!\!$ \\
			                               & $\!\! I_8: r_3 \!=\! a \!+\! r_2 \!-\! 1 \!\!$ \\
			                               & $\!\! I_9: r_4=\LdInst\ r_3 \!\!$ \\
			\hline
			\multicolumn{2}{|l|}{RMO forbids: $r_1=1,\ r_2=1$} \\
			\multicolumn{2}{|l|}{$r_3=a,\ r_4=0$} \\
			\hline
		\end{tabular}
		\caption{Speculation in RMO} \label{fig: rmo spec}
	\end{minipage}
	\hspace{1pt}
	\begin{minipage}[b]{0.4\columnwidth}
		\small
		\centering
		\begin{tabular}{|l|l|}
			\hline
			Proc. P1 & {Proc. P2$\!\!$} \\ 
			\hline
			$\!\! I_1: r_1 = \LdInst\ a \!\!$ & $\!\! I_3: \StInst\ a\ 1 \!\!$ \\ 
			$\!\! I_2: r_2 = \LdInst\ a \!\!$ & \\ 
			\hline
			\multicolumn{2}{|l|}{RMO allows: $r_1 = 1,$} \\ 
			\multicolumn{2}{|l|}{$r_2 = 0$} \\ 
			\hline
		\end{tabular}
		\caption{CoRR in RMO} \label{fig: corr rmo}
	\end{minipage}
\end{figure}

Besides the above problem, RMO permits the reordering of two loads to the same address, \ie{} it allows the non-SC behavior in the Coherent Read-Read (CoRR) example in Figure \ref{fig: corr rmo}.
Although this is not wrong, it makes the compilation from C++ \cite{c++n4527} inefficient, because such behavior is forbidden by C++ even for relaxed atomic loads and stores. 
One could fix this problem by adding the following axiom:
\begin{itemize}
	\item \emph{\rmoFixLd}: the program order of two loads to the same address must be preserved in memory order.
\end{itemize}
However this fix rules out an optimization implemented in Power and ARM processors (see Section \ref{sec: wmm dep litmus}).

\section{Defining Memory Models Using I\textsuperscript{2}E} \label{sec: I2E}

\begin{figure}[!htb]
	\centering
	\includegraphics[width=0.35\textwidth]{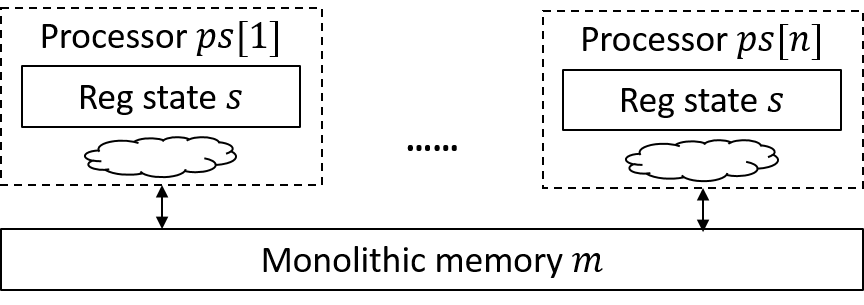}
	\caption{General model structure} \label{fig: gen op model}
\end{figure}

We will model multiprocessor systems as shown in Figure \ref{fig: gen op model} to define memory models. 
The state of the system with $n$ processors is defined as $\langle ps, m \rangle$, where $m$ is an $n$-ported \emph{monolithic} memory which is connected to the $n$ processors. 
$ps[i]$ ($i=1\ldots n$) represents the state of the $i^{th}$ processor. 
Each processor contains a register state $s$, which represents all architectural registers, including both the general purpose registers and special purpose registers, such as PC. 
Cloud represents additional state elements, \eg{} a \emph{store buffer},  that a specific memory model may use in its definition.

\subsection{Abstracting the Instruction Set}

Memory model is always part of the ISA. 
However, we want our definitions of the memory models to be as generic as possible. 
For this reason, we introduce the concept of \emph{decoded instruction set} (DIS).
A \emph{decoded instruction} contains all the information of an instruction after it has been decoded and has read all source registers.
To begin with our DIS has the following three instructions. 
\begin{itemize}
	\item $\langle \NmInst, dst, v \rangle$: instructions that do not access memory, such as ALU or branch instructions.
	It writes the computation result $v$ into destination register $dst$.
	\item $\langle \LdInst, a, dst \rangle$: a load that reads memory address $a$ and updates the destination register $dst$.
	\item $\langle \StInst, a, v \rangle$: a store that writes value $v$ to memory address $a$.
\end{itemize}
Later we will extend the DIS with fence instructions as needed. 
Next we explain how we get to decoded instructions from the source or raw instructions.

\noindent\textbf{Instantaneous Instruction Execution ({I\textsuperscript{2}E}):}
To define memory models we restrict ourselves to \IIE{} models where each instruction is executed instantaneously and the register state of each processor is by definition always up-to-date. Therefore we can define the following two methods on each processor to manipulate the register state $s$:
\begin{itemize}
	\item $\decodeFunc()$: fetches the next raw instruction and returns the corresponding decoded instruction based on the current register state $s$.
	\item $\executeFunc(dIns, ldRes)$: updates the register state $s$ (\eg{} by writing destination registers and changing PC) according to the current decoded instruction $dIns$.
	A $\LdInst$ requires a second argument $ldRes$ which should be the loaded value.
	For other instructions, the second argument can be set to don't care (``$\Hyphen$").
\end{itemize}
\IIE{} cannot describe the semantics of a memory model in which the meaning of an instruction may depend upon a future store.
Therefore all \IIE{} models we discuss do no permit stores to overtake loads.

\subsection{Notations for Operational Semantics}

The operational semantics is a set of \emph{rules} that describe how the state of the abstract machine evolves as execution progresses.
Each rule takes the following form:
\begin{displaymath} 
\frac{predicates\ (based\ on\ the\ current\ state)}{actions\ (to\ modify\ the\ current\ state)}
\end{displaymath} 
If all $predicates$ of a rule are \emph{satisfied} then it can \emph{fire} and atomically update model states according to the specified $actions$.

A predicate is either a \emph{when} statement or a \emph{pattern matching} statement. For example, $\whenFunc(b.\emptyFunc())$ means that the rule requires buffer $b$ to be empty in order to fire. The \emph{pattern matching} statement has the following form:
\begin{displaymath}
pattern = expression
\end{displaymath}
For example, if we want to match the instruction returned by the $\decodeFunc()$ method to be a $\NmInst$ instruction, we can write $\langle \NmInst, dst, v \rangle = ps[i].\decodeFunc()$.
Free variables $dst$ and $v$ will be assigned to appropriate values if the matching is successful.
The matching identifier always begins with a capital letter, \eg{} $\NmInst$, $\LdInst$, $\StInst$, \etc{}

\begin{figure}[!htb]
	\centering
	\subfloat[SC\label{fig: sc op model}]{\includegraphics[width=0.309\columnwidth]{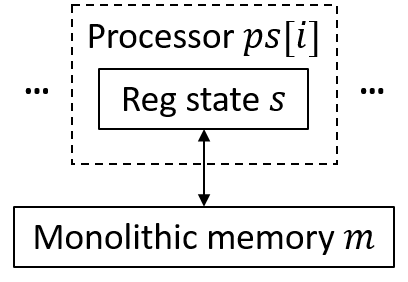}}
	\subfloat[TSO/PSO\label{fig: tso op model}]{\includegraphics[width=0.309\columnwidth]{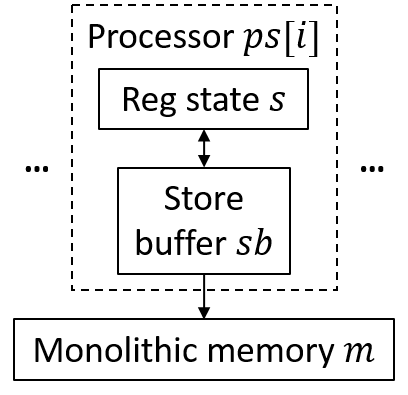}}
	\subfloat[WMM\label{fig: wmm op model}]{\includegraphics[width=0.382\columnwidth]{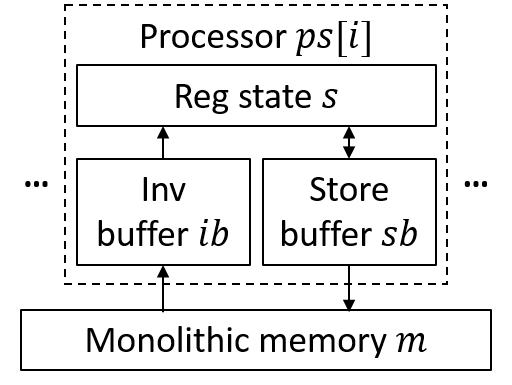}}
	\caption{Structures of \IIE{} models}
\end{figure}

We use ``$\Leftarrow$" to assign a new value to a state, and use semicolon ``;" to separate statements written on the same line.
If multiple rules can fire, then our semantic model selects any one of those rules \emph{non-deterministically} to execute.
The final outcome may depend on the choice of rule selection.
To better understand the notation, we will give \IIE{} descriptions of three well-known memory models: SC, TSO and PSO (see Figures \ref{fig: sc op model} and \ref{fig: tso op model}).

\subsection{SC Model}
As shown in Figure \ref{fig: sc op model}, SC does not require any special buffer.
Figure \ref{fig: sc rule} shows the operational semantics of SC.
The three rules correspond to the \emph{instantaneous execution} of the three types of decoded instructions.
In each rule, the $\decodeFunc()$ method first fetches and decodes a new instruction, and then the instruction is immediately executed and committed.
Loads and stores in SC directly access the monolithic memory.

\begin{figure}[!htb]
	\centering
	\begin{boxedminipage}{\columnwidth}
		\small
		\textbf{\scNmRule{} rule} ($\NmInst$ execution). \reduceRuleSpace
		\begin{displaymath}
		\frac{
			\langle \NmInst, dst, v \rangle = ps[i].\decodeFunc()
		}{
			ps[i].\executeFunc(\langle \NmInst, dst, v \rangle, \Hyphen)
		}
		\end{displaymath} \reduceRuleEndSpace
		
		\textbf{\scLdRule{} rule} ($\LdInst$ execution). \reduceRuleSpace
		\begin{displaymath}
		\frac{
			\langle \LdInst, a, dst \rangle = ps[i].\decodeFunc()
		}{
			ps[i].\executeFunc(\langle \LdInst, a, dst \rangle, m[a])
		}
		\end{displaymath} \reduceRuleEndSpace
		
		\textbf{\scStRule{} rule} ($\StInst$ execution). \reduceRuleSpace
		\begin{displaymath}
		\frac{
			\langle \StInst, a, v \rangle = ps[i].\decodeFunc()
		}{
			ps[i].\executeFunc(\langle \StInst, a, v \rangle, \Hyphen);\ m[a] \Leftarrow v
		}
		\end{displaymath}
	\end{boxedminipage}
	\caption{SC operational semantics} \label{fig: sc rule}
\end{figure}

\subsection{TSO Model}

Figure \ref{fig: tso op model} shows the states and structure of TSO operational model proposed in \cite{owens2009better,sewell2010x86}.
In addition to register state $s$, each processor now contains a store buffer $sb$.
$sb$ is an unbounded buffer of $\langle \mathrm{address, value} \rangle$ pairs, each representing a pending store.
The following methods are defined on $sb$:
\begin{itemize}
	\item $\enqFunc(a, v)$: enqueues the $\langle \mathrm{address, value} \rangle$ pair $\langle a, v \rangle$ into $sb$.
	\item $\deqFunc()$: deletes the oldest store from sb, and returns its $\langle \mathrm{address, value} \rangle$ pair.
	\item $\emptyFunc()$: returns $\True$ when $sb$ is empty.
	\item $\existFunc(a)$: returns $\True$ if address $a$ is present in $sb$.
	\item $\getYoungestFunc(a)$: returns the store data of the youngest store to address $a$ in $sb$.
\end{itemize}
For methods that update the state and return a value, we use "$\leftarrow$" as in $\langle a,v \rangle \leftarrow sb.\deqFunc()$, which assigns the return value of $\deqFunc()$ to pair $\langle a,v \rangle$.

In order to enforce instruction ordering in accessing the newly added store buffer, we extend our instruction set with the memory fence instruction called $\ComInst$ which flushes the local store buffer.

Figure \ref{fig: tso rule} shows the operational semantics of TSO.
The first four rules are instantaneous execution of four types of decoded instructions, while the fifth rule handles the interaction between the store buffer and monolithic memory.
According to \tsoLdRule{}, $\LdInst\ a$ first tries to read the youngest store to address $a$ in the local $sb$, and if $sb$ does not contain $a$ then it reads the monolithic memory. 
Buffering stores in $sb$ essentially allows a load to overtake a store.
The $\ComInst$ fence blocks until older stores are flushed from the store buffer.
The buffer will eventually get empty as the consequence of repeated execution of \tsoDeqSbRule. 

\begin{figure}[!htb]
	\centering
	\begin{boxedminipage}{\columnwidth}
		\small
		\textbf{\tsoNmRule{} rule} ($\NmInst$ execution). \reduceRuleSpace
		\begin{displaymath}
		\frac{
			\langle \NmInst, dst, v \rangle = ps[i].\decodeFunc()
		}{
			ps[i].\executeFunc(\langle \NmInst, dst, v \rangle, \Hyphen)
		}
		\end{displaymath} \reduceRuleEndSpace
		
		\textbf{\tsoLdRule{} rule} ($\LdInst$ execution). \reduceRuleSpace
		\begin{displaymath}
		\frac{
			\begin{array}{c}
				\langle \LdInst, a, dst \rangle = ps[i].\decodeFunc() \\
				v = \ifFunc\ ps[i].sb.\existFunc(a)\ \thenFunc\ ps[i].sb.\getYoungestFunc(a)\ \elseFunc\ m[a] \\
			\end{array}
		}{
			ps[i].\executeFunc(\langle \LdInst, a, dst \rangle, v)
		}
		\end{displaymath} \reduceRuleEndSpace
		
		\textbf{\tsoStRule{} rule} ($\StInst$ execution). \reduceRuleSpace
		\begin{displaymath}
		\frac{
			\langle \StInst, a, v \rangle = ps[i].\decodeFunc()
		}{
			ps[i].\executeFunc(\langle \StInst, a, v \rangle, \Hyphen);\ ps[i].sb.\enqFunc(a, v)
		}
		\end{displaymath} \reduceRuleEndSpace
		
		\textbf{\tsoComRule{} rule} ($\ComInst$ execution). \reduceRuleSpace
		\begin{displaymath}
		\frac{
			\langle \ComInst \rangle = ps[i].\decodeFunc();\ \whenFunc(ps[i].sb.\emptyFunc())
		}{
			ps[i].\executeFunc(\langle \ComInst \rangle, \Hyphen)
		}
		\end{displaymath} \reduceRuleEndSpace
		
		\textbf{\tsoDeqSbRule{} rule} (dequeue TSO store buffer). \reduceRuleSpace
		\begin{displaymath}
		\frac{
			\whenFunc(\neg ps[i].sb.\emptyFunc())
		}{
			\langle a,v \rangle \leftarrow ps[i].sb.\deqFunc();\ m[a] \Leftarrow v
		}
		\end{displaymath}

		\textbf{\psoDeqSbRule{} rule} (dequeue PSO store buffer). \reduceRuleSpace
		\begin{displaymath}
			\frac{
				a = ps[i].sb.\getAnyAddr();\ \whenFunc(a \neq \epsilon)
			}{
				v \leftarrow ps[i].sb.\removeOldestFunc(a);\ m[a] \Leftarrow v 
			}
		\end{displaymath}
	\end{boxedminipage}
	\caption{TSO/PSO operational semantics} \label{fig: tso rule}
\end{figure}

\subsubsection{PSO: Enabling Store-Store reordering}

We extend TSO to PSO by allowing $sb$ to commit the oldest store of any address to the monolithic memory, \ie{} replacing the \tsoDeqSbRule{} rule by the \psoDeqSbRule{} rule as shown in Figure \ref{fig: tso rule}.
We use the following two methods, instead of $\deqFunc()$, to delete entries from $sb$.
\begin{itemize}
	\item $\getAnyAddr()$: returns any store address present in $sb$; or returns $\epsilon$ if $sb$ is empty.
	\item $\removeOldestFunc(a)$: deletes the oldest store to address $a$ from $sb$, and returns its store data.
\end{itemize}
Thus we can dequeue stores for different addresses from the same store buffer out of order, \ie{} reorder stores.

\section{WMM Model} \label{sec: WMM}

WMM allows Load-Load reordering in addition to the reorderings allowed by PSO.
Since a reordered load may read a stale value, we introduce a conceptual device called \emph{invalidation buffer}, $ib$, to each processor (see Figure \ref{fig: wmm op model}).
$ib$ is an unbounded buffer of $\langle \mathrm{address, value} \rangle$ pairs, each representing a stale memory value for an address that can be observed by the processor. 
A stale value enters $ib$ when some store buffer pushes a value to the monolithic memory.
Similar to $\ComInst$ fences for store buffers, we need a $\RecInst$ memory fence to flush the local invalidation buffer.

The following methods are defined on $ib$:
\begin{itemize}
	\item $\insertFunc(a,v)$: inserts $\langle \mathrm{address, value} \rangle$ pair $\langle a, v \rangle$ into $ib$.
	\item $\getRandAndRemoveFunc(a)$: returns a random value $v$ for address $a$ present in $ib$, and removes all values for $a$, which are inserted into $ib$ before $v$, from $ib$.
	\item $\clearFunc()$: removes all contents from $ib$ to make it empty.
	\item $\removeAddrFunc(a)$: removes all (stale) values for address $a$ from $ib$.
\end{itemize}
$\clearFunc$ and $\removeAddrFunc$ are called when ordering needs to be enforced, \ie{} preventing younger loads from reading stale values.

\subsection{Operational Semantics of WMM}

Figure \ref{fig: wmm rule} shows the operational semantics of WMM.
The first 7 rules are the instantaneous execution of decoded instructions, while the \wmmDeqSbRule{} rule removes the oldest store for any address (say $a$) from $sb$ and commits it to the monolithic memory. 
\wmmDeqSbRule{} also inserts the original memory value into the $ib$ of all other processors to allow $\LdInst\ a$ in these processors to effectively get reordered with older instructions. 
However, this insertion in $ib$ should not be done if the corresponding $sb$ on that processor already has a store to $a$. 
This restriction is important, because if a processor has address $a$ in its $sb$, then it can never see stale values for $a$. 
For the same reason, when a $\StInst\ a\ v$ is inserted into $sb$, we remove all values for $a$ from the $ib$ of the same processor.

\begin{figure}[!htb]
	\centering
	\begin{boxedminipage}{\columnwidth}
		\small
		\textbf{\wmmNmRule{} rule} ($\NmInst$ execution). \reduceRuleSpace
		\begin{displaymath}
		\frac{
			\langle \NmInst, dst, v \rangle = ps[i].\decodeFunc()
		}{
			ps[i].\executeFunc(\langle \NmInst, dst, v \rangle, \Hyphen)
		}
		\end{displaymath} \reduceRuleEndSpace
		
		\textbf{\wmmLdSbRule{} rule} ($\LdInst$ execution: bypass from store). \reduceRuleSpace
		\begin{displaymath}
		\frac{
			\langle \LdInst, a, dst \rangle = ps[i].\decodeFunc();\ \whenFunc(ps[i].sb.\existFunc(a))
		}{
			ps[i].\executeFunc(\langle \LdInst, a, dst \rangle, ps[i].sb.\getYoungestFunc(a))
		}
		\end{displaymath} \reduceRuleEndSpace
		
		\textbf{\wmmLdMemRule{} rule} ($\LdInst$ execution: read memory). \reduceRuleSpace
		\begin{displaymath}
		\frac{
			\langle \LdInst, a, dst \rangle = ps[i].\decodeFunc();\ \whenFunc(\neg ps[i].sb.\existFunc(a))
		}{
			ps[i].\executeFunc(\langle \LdInst, a, dst \rangle, m[a]);\ ps[i].ib.\removeAddrFunc(a)
		}
		\end{displaymath} \reduceRuleEndSpace
		
		\textbf{\wmmLdIbRule{} rule} ($\LdInst$ execution: read stale value). \reduceRuleSpace
		\begin{displaymath}
		\frac{
			\begin{array}{c}
			\langle \LdInst, a, dst \rangle = ps[i].\decodeFunc() \\
			\whenFunc(\neg ps[i].sb.\existFunc(a)\ \wedge\ ps[i].ib.\existFunc(a)) \\
			\end{array}
		}{
			v \leftarrow ps[i].ib.\getRandAndRemoveFunc(a);\ ps[i].\executeFunc(\langle \LdInst, a, dst \rangle, v)
		}
		\end{displaymath} \reduceRuleEndSpace
		
		\textbf{\wmmStRule{} rule} ($\StInst$ execution). \reduceRuleSpace
		\begin{displaymath}
		\frac{
			\langle \StInst, a, v \rangle = ps[i].\decodeFunc()
		}{
			ps[i].\executeFunc(\langle \StInst, a, v \rangle, \Hyphen);\ ps[i].sb.\enqFunc(a, v);\ ps[i].ib.\removeAddrFunc(a)
		}
		\end{displaymath} \reduceRuleEndSpace
		
		\textbf{\wmmComRule{} rule} ($\ComInst$ execution). \reduceRuleSpace
		\begin{displaymath}
		\frac{
			\langle \ComInst \rangle = ps[i].\decodeFunc();\ \whenFunc(ps[i].sb.\emptyFunc())
		}{
			ps[i].\executeFunc(\langle \ComInst \rangle, \Hyphen)
		}
		\end{displaymath} \reduceRuleEndSpace
		
		\textbf{\wmmRecRule{} rule} ($\RecInst$ execution). \reduceRuleSpace
		\begin{displaymath}
		\frac{
			\langle \RecInst \rangle = ps[i].\decodeFunc()
		}{
			ps[i].ib.\clearFunc();\ ps[i].\executeFunc(\langle \RecInst \rangle, \Hyphen)
		}
		\end{displaymath} \reduceRuleEndSpace
		
		\textbf{\wmmDeqSbRule{} rule} (dequeue store buffer). \reduceRuleSpace
		\begin{displaymath}
		\frac{
			\begin{array}{c}
			a = ps[i].sb.\getAnyAddr();\ old = m[a];\ \whenFunc(a \neq \epsilon) \\
			\end{array}
		}{
			\begin{array}{c}
			v \leftarrow ps[i].sb.\removeOldestFunc(a);\ m[a] \Leftarrow v \\
			\forall j \neq i.\ \ifFunc\ \neg ps[j].sb.\existFunc(a)\ \thenFunc\ ps[j].ib.\insertFunc(a,old) \\
			\end{array}
		}
		\end{displaymath}
	\end{boxedminipage}
	\caption{WMM operational semantics} \label{fig: wmm rule}
\end{figure}

Load execution rules in Figure \ref{fig: wmm rule} correspond to three places from where a load can get its value.
\wmmLdSbRule{} executes $\LdInst\ a$ by reading from $sb$.
If address $a$ is not found in $sb$, then the load can read from the monolithic memory (\wmmLdMemRule).
However, in order to allow the load to read a stale value (to model load reordering), \wmmLdIbRule{} gets the value from $ib$.
The model allows {non-deterministic} choice in the selection of \wmmLdMemRule{} and \wmmLdIbRule{}.
To make this idea work, \wmmLdMemRule{} has to remove all values for $a$ from $ib$,  because these values are staler than the value in memory.
Similarly, \wmmLdIbRule{} removes all the values for $a$, which are staler than the one read, from $ib$.

\noindent\textbf{Synchronization instructions:}
Atomic read-modify-write ($\RMWInst$) instructions can also be included in WMM.
$\RMWInst$ should directly operate on the monolithic memory, so the rule to execute $\RMWInst$ is simply the combination of \wmmLdMemRule, \wmmStRule{} and \wmmDeqSbRule.
One could also extend WMM to incorporate the load-linked/store-conditional pair in a similar way.

\subsection{Litmus Tests for WMM} \label{sec: wmm litmus test}

WMM executes instructions instantaneously and in order, but because of store buffers ($sb$) and invalidation buffers ($ib$), a processor can see the effect of loads and stores on some other processor in a different order than the program order on that processor.
We explain the reorderings permitted and forbidden by the definition of WMM using well-known examples.

\noindent\textbf{Fences for mutual exclusion:}
Figure \ref{fig: dekker wmm} shows the kernel of Dekker's algorithm in WMM, which guarantees mutual exclusion by ensuring registers $r_1$ and $r_2$ cannot both be zero at the end.
Fences $I_2,I_3,I_6,I_7$ are needed to keep this invariant.
Suppose we remove a $\RecInst$ fence $I_3$. 
Consider the scenario that all instructions on P2 execute first and $I_8$ gets 0.
After that, all instruction on P1 execute and $I_4$ reads the stale value 0 from $ib$.
It is as if $I_4$  overtakes $I_1$ and $I_2$.
If we alternatively remove a $\ComInst$ fence $I_2$, consider the case that all instructions on P1 execute first and $I_4$ gets 0.
Let $I_1$ keep staying in the $sb$ of P1.
Meanwhile, we execute all instructions on P2, so $I_8$ will also get 0.

\begin{figure}[!htb]
	\centering
	\begin{minipage}[b]{0.47\columnwidth}
		\centering
		\small
		\begin{tabular}{|l|l|}
			\hline
			{Proc. P1} & {Proc. P2} \\ 
			\hline
			$\!\! I_1: \StInst\ a\ 1 \!\!$    & $\!\! I_5: \StInst\ b\ 1 \!\!$ \\
			$\!\! I_2: \ComInst \!\!$         & $\!\! I_6: \ComInst \!\!$ \\
			$\!\! I_3: \RecInst \!\!$         & $\!\! I_7: \RecInst \!\!$ \\
			$\!\! I_4: r_1 = \LdInst\ b \!\!$ & $\!\! I_8: r_2 = \LdInst\ a \!\!$ \\ \hline
			\multicolumn{2}{|l|}{\hspace{-3pt}WMM forbids: $r_1 = 0, r_2 = 0 \!\!$} \\ \hline
		\end{tabular}
		\caption{Dekker's algorithm in WMM} \label{fig: dekker wmm}
	\end{minipage}
	\hspace{1pt}
	\begin{minipage}[b]{0.47\columnwidth}
		\centering
		\small
		\begin{tabular}{|l|l|}
			\hline
			{Proc. P1} & {Proc. P2} \\ 
			\hline
			$\!\! I_1: \StInst\ a\ 42$  & $\!\! I_4: r_1 = \LdInst\ f \!\!$ \\
			$\!\! I_2: \ComInst$        & $\!\! I_5: \RecInst \!\!$ \\
			$\!\! I_3: \StInst\ f\ 1$   & $\!\! I_6: r_2 = \LdInst\ a \!\!$ \\ 
			\hline
			\multicolumn{2}{|l|}{\hspace{-3pt}WMM forbids: $r_1=1, r_2 = 0 \!\!$} \\ 
			\hline
		\end{tabular}
		\caption{Message passing in WMM} \label{fig: msg pass wmm}
	\end{minipage}
\end{figure}

\noindent\textbf{Fences for message passing:}
Figure \ref{fig: msg pass wmm} shows a way of inter-processor communication in WMM.
P1 writes data 42 to addresses $a$, and then signals P2 by setting a flag at address $f$ to 1.
P2 sees the new value of $f$ and then reads the data.
Fences $I_2$ and $I_5$ are needed to ensure that the data is correctly passed to P2.
Without the $\ComInst$ fence $I_2$, the data 42 may stay in the $sb$ of P1 even after the flag has been set in the monolithic memory, and P2 may not see the new data.
It is as if the two stores on P1 are reordered.
Without the $\RecInst$ fence $I_5$, P2 could see the stale value 0 from $ib$.
It is as if the two loads on P2 are reordered.

\noindent\textbf{No thin-air read:}
The Thin-air Read behavior in Figure \ref{fig: rc thin air} is impossible in WMM because of \IIE.

\noindent\textbf{SC for a single address}:
WMM maintains SC for all accesses to a single memory location.
This is because both the store buffer and invalidation buffer have the FIFO property for values of the same address.
Therefore WMM does not appear to reorder loads to the same address, or stores to the same address.
For example, the non-SC behavior in the CoRR example in Figure \ref{fig: corr rmo} is forbidden by WMM. 

\noindent\textbf{SC for well-synchronized programs:}
The critical sections in well-synchronized programs are all protected by locks.
To maintain SC behaviors for such programs in WMM, we only need to add a $\RecInst$ after acquiring the lock and a $\ComInst$ before releasing the lock.

In summary, WMM can reorder stores to different addresses, and allows a load to overtake other loads (to different addresses), stores and $\ComInst$ fences.
WMM disallows a load to overtake any $\RecInst$ fence.

\subsection{Compiling C++ to WMM} \label{sec: wmm c++}

C++ primitives \cite{c++n4527} can be mapped to WMM instructions in an efficient way as shown in Table \ref{tab: c++ map}.
For the purpose of comparison, we also include a mapping to Power which has been proven correct~\cite{batty2012clarifying}.

\begin{table}[!htb]
	\centering
	\small
	\begin{tabular}{|l|l|l|}
		\hline
		C++ operations & WMM instructions & Power instructions \\
		\hline
		Non-atomic Load  & $\LdInst$             & $\LdInst$ \\ \hline
		Load Relaxed     & $\LdInst$             & $\LdInst$ \\ \hline
		Load Consume     & $\LdInst;\ \RecInst$  & $\LdInst$ \\ \hline
		Load Acquire     & $\LdInst;\ \RecInst$  & $\LdInst;\ \cmpInst;\ \bcInst;\ \isyncInst$ \\ \hline
		Load SC          & $\ComInst;\ \RecInst;$ & $\syncInst;\ \LdInst;\ \cmpInst;$ \\ 
		                 & $\LdInst;\ \RecInst$  & $\bcInst;\ \isyncInst$ \\ \hline
		Non-atomic Store & $\StInst$             & $\StInst$ \\ \hline
		Store Relaxed    & $\StInst$             & $\StInst$ \\ \hline
		Store Release    & $\ComInst;\ \StInst$  & $\lwsyncInst;\ \StInst \!\!$ \\ \hline
		Store SC         & $\ComInst;\ \StInst$  & $\syncInst;\ \StInst$ \\
		\hline
	\end{tabular}
	\caption{Mapping C++ to WMM and Power} \label{tab: c++ map}
\end{table}

The $\ComInst;\RecInst$ sequence in WMM is the same as a $\syncInst$  in Power, and $\ComInst$ is similar to $\lwsyncInst$.
The $\cmpInst;\bcInst;\isyncInst$ sequence in Power can be viewed as a Load-Load fence, so it is similar to a $\RecInst$ fence in WMM.
In case of Store SC in C++, WMM uses a $\ComInst$ while Power uses a $\syncInst$, so WMM effectively saves one $\RecInst$.
On the other hand, Power does not need any fence for Load Consume in C++, while WMM requires a $\RecInst$.
Thus it is difficult to say whether one is more efficient than the other.

\subsection{WMM Implementation} \label{sec: wmm implement}

WMM can be implemented using modern OOO multiprocessors without any change in the microarchitecture, and even the most aggressive optimizations in ROB and cache cannot step beyond WMM.
To demonstrate this, we first show how general OOO microarchitecture is abstracted by the WMM model, and then we discuss detailed hardware optimizations.

\subsubsection{Correspondence of OOO to WMM}

Let us consider an OOO microarchitecture (referred below simply as OOO) with ROB, store buffer and a coherent write-back cache hierarchy.
In OOO, instructions in ROB are committed in order, loads can be issued as soon as its address is known, and a store is enqueued into the store buffer only when the store commits (\ie{} the entries in a store buffer cannot be killed).
We show how WMM captures the behaviors of OOO by giving a correspondence between OOO and WMM.

\noindent\textbf{Cache hierarchy:}
Although the write-back cache hierarchy may process many requests simultaneously and out of order, every request in the cache hierarchy completes by reading or writing an L1 cache line when it has sufficient permissions and the coherent memory value. 
The monolithic memory in WMM contains only such coherent values, and thus each access to L1 corresponds to an access to the monolithic memory in WMM.
This monolithic memory abstraction of a coherent cache hierarchy has been proven by Vijayaraghavan \etal{} \cite{Vijayaraghavan2015}.

\noindent\textbf{Store buffer:}
The state of the store buffer in OOO is represented by the $sb$ in WMM. 
Entry into the store buffer when a store commits in OOO corresponds to the \wmmStRule{} rule.
In OOO, the store buffer only issues the oldest store for some address to memory.
The store is removed from the store buffer when the store updates L1.
The removal from the store buffer exactly corresponds to the \wmmDeqSbRule{} rule. 

\noindent\textbf{ROB and eager loads:}
The commit of each instruction from ROB corresponds to the WMM rule that executes that instruction, and therefore the architectural state in both WMM and OOO must match at the time of commit.
Early execution of a load $L$ to address $a$ with a return value $v$ in OOO can be understood by considering where $\langle a,v \rangle$ resides in OOO when $L$ \emph{commits}.  
The \wmmLdSbRule{} and the \wmmLdMemRule{} rules cover the cases that $\langle a,v \rangle$ is, respectively, in the store buffer or the cache hierarchy when $L$ commits. 
Otherwise $\langle a,v \rangle$ is no longer present in OOO at the time of load commit and must have been overwritten in memory. 
This case corresponds to using the \wmmDeqSbRule{} rule to inserts $\langle a,v \rangle$ into $ib$, and then using the \wmmLdIbRule{} rule to read $v$ from $ib$.

\noindent\textbf{Fences:}
Fences never go into the store buffer or memory.
In OOO, $\ComInst$ can commit from ROB only when the local store buffer is empty. 
$\RecInst$ plays a different role; it stalls all younger loads unless the load can bypass from a store which is younger than the fence in ROB.
The stall prevents younger loads from reading values that have become stale when the $\RecInst$ commits.
This corresponds to clearing $ib$ in WMM.

In general, we can give a WMM execution for any OOO execution following the above correspondence.
Each time the OOO execution commits an instruction $I$ from ROB or removes a store $S$ from store buffer, the coherent memory state, store buffers, and results of committed instructions in OOO are exactly the same as those in WMM when the WMM execution executes $I$ or dequeues $S$ from $sb$, respectively.

\subsubsection{Aggressive Optimizations}

\noindent\textbf{Speculation:}
WMM does not enforce any dependency ordering, so the implementation can do all kinds of speculations, such as branch prediction, memory dependency prediction \cite{chrysos1998memory}, and even load-value prediction \cite{lipasti1996value,ghandour2010potential,perais2014eole,perais2014practical}.
As a result, a load $L$ can be issued as soon as we know its load address, which could even be computed from a predicted value.
When all predictions related to $L$ turn out to be correct, there is no need to check whether the value that $L$ got earlier has become stale, because getting a stale value is captured by reading $ib$ in WMM.

Consider the behavior in Figure \ref{fig: wmm mem dep pred}.
In an implementation with memory dependency prediction, P2 can predict that the store address of $I_5$ is not $a$, and execute $I_6$ early to get value 0.
WMM allows this behavior because $I_6$ can read 0 from $ib$.
Next consider the behavior in Figure \ref{fig: wmm val pred}.
In an implementation with load-value prediction, P2 can predict the result of $I_4$ to be $a$ and execute $I_5$ early to get value 0.
When $I_4$ returns from memory later with value $a$, the prediction on $I_4$ turns out to be correct and the result of $I_5$ can be kept.
WMM also allows this behavior because $I_5$ can read 0 from $ib$.
Note that the behavior in Figure \ref{fig: wmm val pred} is disallowed by RMO, Power and ARM if we change $I_2$ to $\MBInst$ (RMO fence), $\lwsyncInst$ (Power fence) or $\dmbInst$ (ARM fence).
Thus, load-value prediction cannot be directly used in RMO, Power and ARM processors.

\begin{figure}[!htb]
	\begin{minipage}[b]{0.5\columnwidth}
		\centering
		\small
		\begin{tabular}{|l|l|}
			\hline
			Proc. P1 & Proc. P2 \\
			\hline
			$\!\! I_1: \StInst\ a\ 1 \!\!$ & $\!\! I_4: r_1 = \LdInst\ b \!\!$ \\
			$\!\! I_2: \ComInst \!\!$      & $\!\! I_5: \StInst\ (r_1 \!+\! c \!-\! 1)\ 1 \!\!$ \\
			$\!\! I_3: \StInst\ b\ 1 \!\!$ & $\!\! I_6: r_2 = \LdInst\ a \!\!$ \\
			\hline
			\multicolumn{2}{|l|}{WMM allows: $r_1=1,r_2=0$} \\
			\hline
		\end{tabular}
		\caption{Memory dependency prediction} \label{fig: wmm mem dep pred}
	\end{minipage}
	\hspace{1pt}
	\begin{minipage}[b]{0.45\columnwidth}
		\centering
		\small
		\begin{tabular}{|l|l|}
			\hline
			Proc. P1 & Proc. P2 \\
			\hline
			$\!\! I_1: \StInst\ a\ 1 \!\!$ & $\!\! I_4: r_1 = \LdInst\ b \!\!$ \\
			$\!\! I_2: \ComInst \!\!$      & $\!\! I_5: r_2 = \LdInst\ r_1 \!\!$ \\
			$\!\! I_3: \StInst\ b\ a \!\!$ & \\
			\hline
			\multicolumn{2}{|l|}{\hspace{-4pt}WMM allows: $r_1=a,r_2=0 \!\!$} \\
			\hline
		\end{tabular}
		\caption{Load-value prediction} \label{fig: wmm val pred}
	\end{minipage}
\end{figure}

The only restriction on issuing loads comes from the fact that WMM does not appear to reorder loads for the same address.
The implementation could execute such two loads out-of-order, but it must ensure that the return values of the two loads are from the same store.
(Power and ARM have the same restriction \cite{sarkar2011understanding,flur2016modelling}).

\noindent\textbf{Coherence optimization:}
A possible coherence optimization, which we refer to as \emph{delayed-invalidation}, is to delay the processing of invalidation requests.
Suppose the local cache $C$ of processor $Pi$ holds a cache line for address $a$ in the shared state.
When $C$ receives an invalidation request for $a$ from its parent, $C$ could respond without truly evicting the line, thus letting later
loads to read this stale line.
However, the stale line must be evicted when $C$ processes a store to $a$ or before $Pi$ commits a $\RecInst$.
A load that reads the stale line is effectively executed early, and the behavior is captured by reading $ib$ in WMM.
This optimization may violate all dependency orderings, \eg{} the behavior in Figure \ref{fig: wmm val pred}.
However, if no cache miss can be processed before the eviction of the stale line, then this optimization does not break any dependency ordering or affect the memory model.

Since the stores in WMM are multi-copy atomic (\eg{}, WMM disallows the behavior in Figure \ref{fig: rc no ssb}), we have demonstrated that even this coherence optimization is not tied to non-atomic multi-copy stores. 

Since WMM allows common coherence optimizations and all speculations in ROB, the performance of its implementation should not be worse than that of any other weak model (\eg{} Power and ARM).
Furthermore, since Power and ARM cannot directly use load-value prediction, WMM implementation may even have higher performance than Power and ARM.

\section{Modeling Data Dependency} \label{sec: data dep}

Figure \ref{fig: wmm val pred} shows a behavior permitted by WMM but which is not possible unless hardware does load-value prediction or delayed-invalidation optimization. 
This behavior can be prevented by inserting a $\RecInst$ fence between $I_4$ and $I_5$. 
However, the fence may cause performance loss because it would prevent the execution of loads that follow $I_5$ but do not depend on $I_4$.
This is an unnecessary cost because commercial microprocessors do not use value prediction yet, and the delayed-invalidation optimization can be adapted to not affect the memory model.  
To avoid these extra $\RecInst$ fences, we need a memory model that precisely captures the data-dependency ordering enforced in hardware.
As we have seen, the axioms of RMO restrict hardware too much, while Power and ARM explicitly models ROB operations.
None of these solutions are satisfactory, so we present \emph{\wmmDep} which uses \emph{timestamps} to exclude exactly those behaviors that violate data-dependency ordering from WMM.

\subsection{Enforcing Data Dependency with Timestamps} \label{sec: wmm dep intuition}

We derive our intuition for timestamps by observing how an OOO processor without load-value prediction and delayed-invalidation optimization works. 
We refer to such a processor as \emph{\oooNVP}.

In Figure \ref{fig: wmm val pred}, assume instruction $I_k$ ($k=1\ldots 5$) gets its result or writes memory at time $t_k$ in \oooNVP.
Then $t_5 \geq t_4$ because the result of $I_4$ is a source operand of $I_5$ (\ie{} the load address).
Since $I_4$ reads the value of $I_3$ from memory, $t_4 \geq t_3$, and thus $t_5\geq t_3\geq t_1$.
As we can see, the time ordering reflects enforcement of data dependencies.
Thus, a natural way to extend WMM to \wmmDep{} is to attach a timestamp to each value, which will, in turn, impose additional constraints on rule firing in WMM.
We first explain how to extend WMM to \wmmDep{} without considering \oooNVP, and then show the correspondence between \wmmDep{} and \oooNVP.

\subsubsection{Adding Timestamps to WMM} \label{sec: add timestamp}

Let us assume there is a global clock which is incremented every time a store writes memory.
We attach a timestamp to each value in WMM, \ie{} an architecture register value, the $\langle \mathrm{address,value} \rangle$ pair of a store, and a monolithic memory value.
The timestamp represents when the value is created.
Consider an instruction $r_3=r_1+r_2$. 
The timestamp of the new value in $r_3$ will be the maximum timestamp of $r_1$ and $r_2$.
Similarly, the timestamp of the $\langle \mathrm{address,value} \rangle$ pair of a store ($\StInst\ a\ v$), \ie{} the \emph{creation} time of the store, is the maximum timestamp of all source operands to compute $\langle a,v \rangle$.
The timestamp of a monolithic memory value is the time when the value becomes visible in memory, \ie{} one plus the time when the value is stored.

Next consider a load $L$ ($\LdInst\ a$) on processor $i$, which reads the value of a store $S$ ($\StInst\ a\ v$).
No matter how WMM executes $L$ (\eg{} by reading $sb$, memory, or $ib$), the timestamp $ts$ of the load value (\ie{} the timestamp for the destination register) is always the maximum of (1) the timestamp $ats$ of the address operand, (2) the time $rts$ when processor $i$ executes the last $\RecInst$ fence, and (3) the time $vts$ when $S$ becomes visible to processor $i$.
Both $ats$ and $rts$ are straightforward.
As for $vts$, if $S$ is from another processor $j$ ($j\neq i$), then $S$ is visible after it writes memory, so $vts$ is timestamp of the monolithic memory value written by $S$.
Otherwise, $S$ is visible to processor $i$ after it is created, so $vts$ is the creation time of $S$.

A constraint for $L$, which we refer to as \emph{stale-timing}, is that $ts$ should not exceed the time $ts_E$ when $S$ is overwritten in memory.
This constraint is only relevant when $L$ reads from $ib$.
In Section \ref{sec: relate ooo-d}, we will explain why this constraint is needed.

To carry out the above timestamp calculus for load $L$ in WMM, we need to associate the monolithic memory $m[a]$ with the creation time of $S$ and the processor that created $S$, when $S$ updates $m[a]$.
When $S$ is overwritten and its $\langle a,v \rangle$ is inserted into $ps[i].ib$, we need to attach the time interval $[vts, ts_E]$ (\ie{} the duration that $S$ is visible to processor $i$) to that $\langle a,v \rangle$ in $ps[i].ib$.

By combining the above timestamp mechanism with the original WMM rules, we have derived \wmmDep.

\subsubsection{Relation Between \wmmDep{} and \oooNVP} \label{sec: relate ooo-d}

The timestamp of each value in \wmmDep{} represents the \emph{earliest} time that the value may become readable in \oooNVP.
For example, the timestamp of a register value in WMM is the earliest time in \oooNVP{}, at which the value can be derived by an instruction.
It should be noted that PC should never be involved in the timestamp mechanism of \wmmDep{}.
This is because instructions can be speculatively fetched in \oooNVP, and the PC of each instruction can always be known in advance.

As for loads, we first make a simplification that a $\RecInst$ in \oooNVP{} stalls all younger loads, \ie{} bypassing from stores younger than the fence is also stalled.
This restriction does not reduce the permitted behaviors in \oooNVP{}.
This is because no younger load can access memory before the $\RecInst$ commits even without the restriction, and the additionally stalled bypassing can be done immediately after the $\RecInst$ commits.

With the above simplification, a load from processor $i$ can get its value in \oooNVP{} only when (1) its address has resolved, (2) all previous $\RecInst$ fences have committed, and (3) the value is visible to processor $i$.
This exactly corresponds to how we compute the timestamp of the load result in \wmmDep{}.
In terms of the time when the value becomes visible in \oooNVP, if the load value is also stored by processor $i$, then the load can bypass from a store in ROB right after the $\langle \mathrm{address, value} \rangle$ of the store is computed; otherwise the load must wait for the value to be written into memory.
This also corresponds to the computation of $vts$ in \wmmDep{}.

Since it is impossible in \oooNVP{} to have a load get its value after the value has been overwritten in memory, the stale-timing constraint in \wmmDep{} is necessary.

\subsection{\wmmDep{} Operational Semantics}

The operational semantics of \wmmDep{} is given in Figure \ref{fig: wmm dep op model}.
We list the things one should remember before reading the rules in the figure.
\begin{itemize}
	\item The global clock name is $gts$ (initialized as 0), which is incremented when the monolithic memory is updated.
	
	\item Each register has a timestamp (initialized as 0) which indicates when the register value was created.
	
	\item Each $sb$ entry $\langle a,v\rangle$ has a timestamp, \ie{} the creation time of the store that made the entry.
	Timestamps are added to the method calls on $sb$ as appropriate.
	
	\item Each monolithic memory location $m[a]$ is a tuple \\ $\langle v, \langle i, sts \rangle, mts \rangle$ (initialized as $\langle 0, \langle \Hyphen, 0 \rangle, 0 \rangle$), in which $v$ is the memory value, $i$ is the processor that writes the value, $sts$ is the creation time of the store that writes the value, and $mts$ is the timestamp of the memory value (\ie{} one plus the time of memory write). 
	
	\item Each $ib$ entry $\langle a,v \rangle$ has a time interval $[ts_L, ts_U]$, in which $ts_L$ is the time when $\langle a,v \rangle$ becomes visible to the processor of $ib$, and $ts_U$ is the time when $\langle a,v \rangle$ is overwritten in memory and gets inserted into $ib$.
	Thus, the $\insertFunc$ method on $ib$ takes the time interval as an additional argument.
	
	\item Each processor $ps[i]$ has a timestamp $rts$ (initialized as 0), which records when the latest $\RecInst$ was executed by $ps[i]$.
\end{itemize}

Some of the timestamp manipulation is done inside the decode and execute methods of each processor $ps[i]$.
Therefore we define the following methods:
\begin{itemize}
	\item $\decodeTSFunc()$: returns a pair $\langle dIns, ts \rangle$, in which $dIns$ is the decoded instruction returned by the original method $\decodeFunc()$, and $ts$ is the maximum timestamp of all source registers (excluding PC) of $dIns$.
	\item $\executeTSFunc(dIns, ldRes, ts)$: first calls the original method $\executeFunc(dIns, ldRes)$, and then writes timestamp $ts$ to the destination register of instruction $dIns$.
\end{itemize}
We also replace the  $\getRandAndRemoveFunc$ method on $ib$ with the following two methods:
\begin{itemize}
	\item $\getRandFunc(a)$: returns the $\langle \mathrm{value, time\ interval} \rangle$ pair of a random stale value for address $a$ in $ib$.
	If $ib$ does not contain any stale value for $a$, $\langle \epsilon, \Hyphen \rangle$ is returned.
	
	\item $\removeOlderFunc(a, ts)$: removes all stale values for address $a$, which are inserted into $ib$ when $gts < ts$, from $ib$.
\end{itemize}
This facilitates the check of the stale-timing constraint.

\begin{figure}[!htb]
	\centering
	\small
	\begin{boxedminipage}{\columnwidth}
		\textbf{\wmmDepNmRule{} rule} ($\NmInst$ execution). \reduceRuleSpace
		\begin{displaymath}
		\frac{
			\langle \langle \NmInst, dst, v \rangle, ts \rangle = ps[i].\decodeTSFunc()
		}{
			ps[i].\executeTSFunc(\langle \NmInst, dst, v \rangle, \Hyphen, ts)
		}
		\end{displaymath} \reduceRuleEndSpace
		
		\textbf{\wmmDepLdSbRule{} rule} ($\LdInst$ execution: bypass from store). \reduceRuleSpace
		\begin{displaymath}
		\frac{
			\begin{array}{c}
			\langle \langle \LdInst, a, dst \rangle, ats \rangle = ps[i].\decodeTSFunc() \\ 
			\whenFunc(ps[i].sb.\existFunc(a));\ \langle v, sts\rangle = ps[i].sb.\getYoungestFunc(a) \\
			\end{array}
		}{
			ps[i].\executeTSFunc(\langle \LdInst, a, dst\rangle, v, \maxFunc(ats, ps[i].rts, sts))
		}
		\end{displaymath} \reduceRuleEndSpace
		
		\textbf{\wmmDepLdMemRule{} rule} ($\LdInst$ execution: read memory). \reduceRuleSpace
		\begin{displaymath}
		\frac{
			\begin{array}{c}
			\langle \langle \LdInst, a, dst \rangle, ats \rangle = ps[i].\decodeTSFunc();\ \whenFunc(\neg ps[i].sb.\existFunc(a)) \\
			\langle v, \langle j, sts \rangle, mts\rangle = m[a];\ vts = (\ifFunc\ i \neq j\ \thenFunc\ mts\ \elseFunc\ sts) \\
			\end{array}
		}{
			\begin{array}{c}
			ps[i].\executeTSFunc(\langle \LdInst, a, dst\rangle, v, \maxFunc(ats, ps[i].rts, vts)) \\ 
			ps[i].ib.\removeAddrFunc(a) \\
			\end{array}
		}
		\end{displaymath} \reduceRuleEndSpace
		
		\textbf{\wmmDepLdIbRule{} rule} ($\LdInst$ execution: read stale value). \reduceRuleSpace
		\begin{displaymath}
		\frac{
			\begin{array}{c}
			\langle \langle \LdInst, a, dst \rangle, ats \rangle = ps[i].\decodeTSFunc() \\ 
			\langle v, [ts_L, ts_U]\rangle = ps[i].ib.\getRandFunc(a) \\
			\whenFunc(\neg ps[i].sb.\existFunc(a)\ \wedge\ v\neq \epsilon\ \wedge\ ats \leq ts_U) \\
			\end{array}
		}{
			\begin{array}{c}
			ps[i].\executeTSFunc(\langle \LdInst, a, dst\rangle, v, \maxFunc(ats, ps[i].rts, ts_L)) \\ 
			ps[i].ib.\removeOlderFunc(a, ts_U) \\
			\end{array}
		}
		\end{displaymath} \reduceRuleEndSpace
		
		\textbf{\wmmDepStRule{} rule} ($\StInst$ execution). \reduceRuleSpace
		\begin{displaymath}
		\frac{
			\langle \langle \StInst, a, v \rangle, ts \rangle = ps[i].\decodeTSFunc()
		}{
			\begin{array}{c}
			ps[i].\executeTSFunc(\langle \StInst, a, v\rangle, \Hyphen, \Hyphen) \\
			ps[i].sb.\enqFunc(a, v, ts);\ ps[i].ib.\removeAddrFunc(a)
			\end{array}
		}
		\end{displaymath} \reduceRuleEndSpace
		
		\textbf{\wmmDepRecRule{} rule} ($\RecInst$ execution). \reduceRuleSpace
		\begin{displaymath}
		\frac{
			\langle \langle \RecInst \rangle, ts \rangle = ps[i].\decodeTSFunc()
		}{
			ps[i].\executeTSFunc(\langle \RecInst \rangle, \Hyphen, \Hyphen);\ ps[i].ib.\clearFunc();\ ps[i].rts \Leftarrow gts
		}
		\end{displaymath} \reduceRuleEndSpace
		
		\textbf{\wmmDepComRule{} rule} ($\ComInst$ execution). \reduceRuleSpace
		\begin{displaymath}
		\frac{
			\langle \langle \ComInst \rangle, ts \rangle = ps[i].\decodeTSFunc();\ \whenFunc(ps[i].sb.\emptyFunc())
		}{
			ps[i].\executeTSFunc(\langle \ComInst \rangle, \Hyphen, \Hyphen)
		}
		\end{displaymath} \reduceRuleEndSpace
		
		\textbf{\wmmDepDeqSbRule{} rule} (dequeue store buffer). \reduceRuleSpace
		\begin{displaymath}
		\hspace{-1pt}\frac{
			\begin{array}{c}
			a = ps[i].sb.\getAnyAddr();\ \langle v', \langle i', sts' \rangle, mts \rangle = m[a] \\
			ts_U = gts;\ \whenFunc(a \neq \epsilon) \\
			\end{array}
		}{
			\hspace{-3pt}\begin{array}{ll}
			\multicolumn{2}{c}{\langle v, sts \rangle \leftarrow ps[i].sb.\removeOldestFunc(a)} \\ 
			\multicolumn{2}{c}{m[a] \Leftarrow \langle v, \langle i, sts \rangle, gts+1 \rangle;\ gts \Leftarrow gts + 1} \\
			\forall j \neq i. & \hspace{-7pt} \letFunc\ ts_L = (\ifFunc\ j \neq i'\ \thenFunc\ mts\ \elseFunc\ sts')\ \inFunc \\ 
			                  & \hspace{-7pt} \ifFunc\ \neg ps[j].sb.\existFunc(a)\ \thenFunc\ ps[j].ib.\insertFunc(a, v', [ts_L, ts_U]) \\
			\end{array} \hspace{-2pt}
		}
		\end{displaymath}
	\end{boxedminipage}
	\caption{\wmmDep{} operational semantics} \label{fig: wmm dep op model}
\end{figure}

In Figure \ref{fig: wmm dep op model}, \wmmDepNmRule{} and \wmmDepStRule{} compute the timestamps of  a $\NmInst$ instruction result and a store $\langle a,v \rangle$ pair from the timestamps of source registers respectively.
\wmmDepRecRule{} updates $ps[i].rts$ with the current time because a $\RecInst$ is executed.
\wmmDepDeqSbRule{} attaches the appropriate time interval to the stale value inserted into $ib$ as described in Section \ref{sec: add timestamp}.

In all three load execution rules (\wmmDepLdSbRule, \wmmDepLdMemRule, and \wmmDepLdIbRule), the timestamp of the load result is no less than the timestamp of the address operand ($ats$) or the latest $\RecInst$ execution time ($ps[i].rts$).
Besides, the timestamp of the load result is also lower-bounded by the beginning time that the value is readable by the processor of the load ($ps[i]$),
In \wmmDepLdSbRule{} and \wmmDepLdIbRule{}, this beginning time (\ie{} $sts$ or $ts_L$) is stored in the $sb$ or $ib$ entry; while in \wmmDepLdMemRule, this beginning time is one of the two times (\ie{} $sts$ and $mts$) stored in the monolithic memory location depending on whether the memory value $v$ is written by $ps[i]$ (\ie{} whether $i$ is equal to $j$).
In \wmmDepLdIbRule, the stale-timing constraint requires that $\maxFunc(ats, ps[i].rts, ts_L)$ (\ie{} the timestamp of the load value) is no greater than $ts_U$ (\ie{} the time when the stale value is overwritten).
Here we only compare $ats$ with $ts_U$, because $ts_L\leq ts_U$ is obvious, and the clearing of $ib$ done by $\RecInst$ fences already ensures $ps[i].rts \leq ts_U$.

\subsection{Litmus Tests for \wmmDep} \label{sec: wmm dep litmus}

\noindent\textbf{Enforcing data dependency:}
First revisit the behavior in Figure \ref{fig: wmm val pred}.
In \wmmDep, the timestamp of the source operand of $I_5$ (\ie{} the result of $I_4$) is 2, while the time interval of the stale value 0 for address $a$ in the $ib$ of P1 is $[0,0]$.
Thus $I_5$ cannot read the stale value 0, and the behavior is forbidden by \wmmDep.
For a similar reason, \wmmDep{} forbids the behavior in Figure \ref{fig: wmm dep st to ld}, in which $I_4$ carries data dependency to $I_7$ transitively.
In particular, $I_6$ reading from $I_5$ forms a data dependency.
This behavior is also impossible in \oooNVP.

\noindent\textbf{Allowing other speculations:}
The behavior in Figure \ref{fig: wmm mem dep pred} is possible on hardware that performs memory dependency speculation.
\wmmDep{} allows this behavior, because the timestamp of the address operand of $I_6$ is 0, and $I_6$ can read the stale value 0 from $ib$.
For a similar reason, \wmmDep{} allows the behavior in Figure \ref{fig: rmo spec} (assuming $I_2$ becomes $\ComInst$), which can be observed on hardware that speculates over control dependency.
As we can see, \wmmDep{} only excludes behaviors that violate data-dependency ordering, while still allowing implementations to speculate on all other dependencies.

\begin{figure}[!htb]
	\centering
	\begin{minipage}[b]{0.44\columnwidth}
		\centering
		\small
		\begin{tabular}{|l|l|}
			\hline
			Proc. P1 & Proc. P2 \\
			\hline
			$\!\! I_1: \StInst\ a\ 1 \!\!$ & $\!\! I_4: r_1=\LdInst\ b \!\!$ \\
			$\!\! I_2: \ComInst \!\!$      & $\!\! I_5: \StInst\ c\ r_1 \!\!$ \\
			$\!\! I_3: \StInst\ b\ a \!\!$ & $\!\! I_6: r_2=\LdInst\ c \!\!$ \\
			                               & $\!\! I_7: r_3 = \LdInst\ r_2\!\!$ \\
			\hline
			\multicolumn{2}{|l|}{\wmmDep{} forbids: $r_1=a,$} \\
			\multicolumn{2}{|l|}{$r_2=a, r_3=0$} \\
			\hline
		\end{tabular}
		\caption{Transitive data dependency} \label{fig: wmm dep st to ld}
	\end{minipage}
	\hspace{1pt}
	\begin{minipage}[b]{0.51\columnwidth}
		\small
		\centering
		\begin{tabular}{|l|l|}
			\hline
			Proc. P1 & Proc. P2 \\
			\hline
			$\!\! I_1: \StInst\ a\ 1 \!\!$ & $\!\! I_4: r_1=\LdInst\ b \!\!$ \\
			$\!\! I_2: \ComInst \!\!$      & $\!\! I_5: r_2 \!=\! r_1 \!+\! c \!-\! 1 \!\!$ \\
			$\!\! I_3: \StInst\ b\ 1 \!\!$ & $\!\! I_6: r_3=\LdInst\ r_2 \!\!$ \\
			                               & $\!\! I_7: r_4=\LdInst\ c \!\!$ \\
			                               & $\!\! I_8: r_5 = r_4 \!+\! a \!\!$ \\
			                               & $\!\! I_9: r_6=\LdInst\ r_5 \!\!$ \\
			\hline
			\multicolumn{2}{|l|}{\hspace{-3pt}\wmmDep{} allows: $r_1=1, r_2=c \!\!$} \\
			\multicolumn{2}{|l|}{$r_3=0, r_4=0, r_5=a, r_6=0$} \\
			\hline
		\end{tabular}
		\caption{RSW in \wmmDep} \label{fig: rsw}
	\end{minipage}
\end{figure}

\noindent\textbf{Necessity of two timestamps in memory:}
In Figure \ref{fig: wmm dep st to ld}, suppose we change $I_5$ to ``$\StInst\ c\ a$", and insert a $\ComInst$ fence between $I_5$ and $I_6$.
Then the behavior will be possible in \oooNVP, because $I_6$ can bypass data from $I_5$, and $I_7$ can execute early to get 0.
(The newly inserted $\ComInst$ fence cannot stop the bypass). 
\wmmDep{} also allows the behavior.
However, if each monolithic memory location in \wmmDep{} only keeps a single timestamp, which is the time when the memory value becomes visible, then $I_6$ must get value $a$ from $m[c]$ with timestamp 3.
Thus $I_7$ cannot read stale value 0, which has time interval $[0,0]$, from $ib$.
This example shows that the two timestamps in each monolithic memory location are indispensable.

\noindent\textbf{Loads to the same address:}
Remember that two loads to the same address can be executed out of order in \oooNVP{} as long as the two loads read from the same store.
\wmmDep{} also captures this subtle optimization.
Consider the Read-from-Same-Write (RSW) program in Figure \ref{fig: rsw}.
The behavior is observable in \oooNVP, because $I_7$ to $I_9$ can be executed before $I_4$ to $I_6$.
It is fine for $I_6$ and $I_7$, which read the same address $c$, to be executed out-of-order, because they both read from the initialization store.
\wmmDep{} allows this behavior, because the timestamp of the address operand of $I_9$ is 0, and $I_9$ can read stale value 0 from $ib$.
(This behavior is also observable on Power and ARM processors \cite{sarkar2011understanding,flur2016modelling}).

In contrast, RMO with the additional \rmoFixLd{} axiom, which disables the reordering of loads to the same address in Section \ref{sec: rmo}, will forbid this behavior (assuming we change $I_2$ to $\MBInst$).
This is because the memory order of $I_4,I_6,I_7,I_9$ must be the same as the program order on P2.
This reveals the disadvantage of the axiomatic definition of RMO.
Maybe adding complicated axioms can disallow the reordering of loads to the same address while capturing this optimization; we certainly have not figured it out.

\subsection{Compiling C++ to \wmmDep}

The mapping from C++ to \wmmDep{} is almost the same as the one for WMM except that \wmmDep{} does not need any fence for Load Consume in C++.
This is because Load Consume leverages data-dependency ordering which is already enforced by \wmmDep.

\section{Modeling Non-Atomic Multi-Copy Stores} \label{sec: non atomic mem}

Unlike a multi-copy atomic store, a non-atomic multi-copy store may become visible to different processors at different times.
This can happen because of shared store buffers or write-through caches.
Even then, all stores for an address can be put in a total order, and the order seen by any processor is consistent with this total order.
We will refer to this total order as the \emph{coherence order} ($\coOrd$) \cite{sarkar2011understanding,mador2012axiomatic}, though in the literature other names, such as modification order \cite{c++n4527}, have also been used.
We can model such stores by introducing a background rule to make copies of a store in a store buffer into other store buffers.
There are quite a few subtleties in doing this properly; \emph{\wmmSSB{}} model is an \IIE{} description of rules to generate such behaviors.

\subsection{Copying From One Store Buffer into Another} \label{sec: st propagate}

We need a mechanism to identify all the copies of a store in various store buffers. 
We, therefore, assign a unique tag $t$ when a store is inserted in the store buffer by a store instruction, and this tag is copied when a store is copied from one store buffer into another.
When it is time to commit a store from the store buffer to the memory, all the copies of this store have to be deleted from all the store buffers which have them. 
A store can be committed only if all its copies are the oldest store for that address in their respective store buffers.

All the stores for an address in a store buffer are kept as a strictly ordered list where the \emph{youngest} store (\ie{} the \emph{largest} in the total order of this list) is the one that entered the store buffer \emph{last}. 
We make sure that all ordered lists are can be combined transitively to form a strict partial order, which has now to be understood in terms of the tags on stores because of the copies.
By the end of the program, this partial order on the stores for an address becomes the coherence order, so we refer to this partial order as the \emph{partial coherence order}.

\begin{figure}[!htb]
	\centering
	\includegraphics[width=0.9\columnwidth]{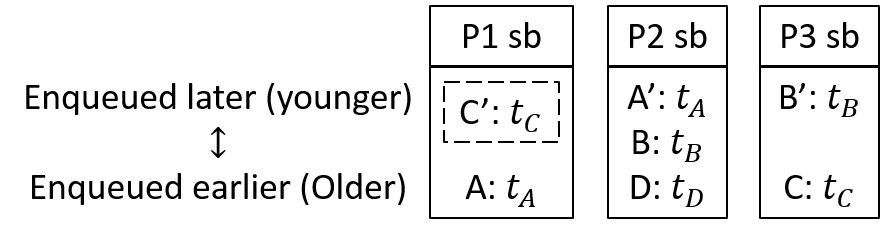}
	\caption{Example states of store buffers by copying stores (primes are copies)} \label{fig: store propagate example}
\end{figure}

Consider the states of store buffers shown in Figure \ref{fig: store propagate example}.
$A$, $B$, $C$ and $D$ are different stores to the same address, and their tags are $t_A$, $t_B$, $t_C$ and $t_D$, respectively.
$A'$ and $B'$ are copies of $A$ and $B$ respectively created by the background copy rule.
Ignoring $C'$, the partial coherence order contains:
\begin{itemize}
\item $t_D \coOrd t_B \coOrd t_A$ ($D$ is older than $B$, and $B$ is older than $A'$ in P2) and 
\item $t_C \coOrd t_B$ ($C$ is older than $B'$ in P3)
\end{itemize}
Notice that $t_D$ and $t_C$ are not related in this partial order.

At this point, if we allowed $C$ in P3 to be copied as $C'$ into P1, we would introduce a new edge $t_A \coOrd t_C$ in the coherence relation, which would break the partial order by introducing the cycle $t_A \coOrd t_C \coOrd t_B \coOrd t_A$.
Therefore copying of $C$ into P1 should not be allowed in this state.
Similarly, copying a store with tag $t_A$ into P1 or P2 should be forbidden because it would immediately create the cycle,  $t_A \coOrd t_A$.
In general, the background copy rule must be constrained so that invariance of the partial coherence order after copying is maintained.

The operational semantics of \wmmSSB{} is defined by adding/replacing three rules to the operational semantics of WMM given in Figure \ref{fig: wmm rule}.
These new rules are shown in Figure \ref{fig: wmm ssb op model}: 
A new background rule \wmmSSBPropSt{} is added to the WMM rules and the \wmmSSBStRule{} and \wmmSSBDeqSbRule{} rules replace the \wmmStRule{} and \wmmDeqSbRule{} rules of WMM, respectively.
Before reading these new rules, one should note the following facts:
\begin{itemize}
	\item The $\decodeFunc$ method now returns $\langle \StInst, a, v, t\rangle$ for a store, in which $t$ is the unique tag assigned to the store.
	Each store buffer entry becomes a tuple $\langle a,v,t \rangle$, in which $t$ is the tag.
	Tags are also introduced into the methods of $sb$ appropriately.
	
	\item The $sb$ now has the following three methods:
	\begin{itemize}
		\item $\hasTagFunc(t)$: returns $\True$ if $sb$ contains a store with tag $t$.
		\item $\getOldestFunc(a)$: returns the $\langle \mathrm{value, tag} \rangle$ pair of the oldest store to address $a$ in $sb$.
		\item $\getRandFunc(a)$: returns the $\langle \mathrm{value, tag} \rangle$ pair of any store to address $a$ present in $sb$.
	\end{itemize}
	If $sb$ does not contain any store to address $a$, $\getOldestFunc(a)$, $\getRandFunc(a)$ and the previously defined $\getYoungestFunc(a)$ methods will all return $\langle \epsilon, \epsilon \rangle$.
	
	\item A new function $\noCycleFunc(a, t, j)$ is defined to check whether the background rule could copy a store with tag $t$ for address $a$ into the $sb$ of processor $j$.
	It returns $\True$ if the partial coherence order among the tags of all stores for address $a$ does not contain any cycle after doing the copy.
\end{itemize}

\begin{figure}[!htb]
	\centering
	\begin{boxedminipage}{\columnwidth}
		\small
		\textbf{\wmmSSBStRule{} rule} ($\StInst$ execution). \reduceRuleSpace
		\begin{displaymath}
		\frac{
			\langle \StInst, a, v, t \rangle = ps[i].\decodeFunc() 
		}{
			\begin{array}{c}
			ps[i].\executeFunc(\langle \StInst, a, v, t \rangle, \Hyphen) \\
			ps[i].sb.\enqFunc(a, v, t);\ ps[i].ib.\removeAddrFunc(a) \\
			\end{array}
		}
		\end{displaymath} \reduceRuleEndSpace
		
		\textbf{\wmmSSBDeqSbRule{} rule} (dequeue store buffer). \reduceRuleSpace
		\begin{displaymath}
		\frac{
			\begin{array}{c}
			a = ps[i].sb.\getAnyAddr();\ old = m[a];\ \langle v,t \rangle = ps[i].sb.\getOldestFunc(a) \\
			\multicolumn{1}{l}{\whenFunc(a \neq \epsilon\ \wedge} \\
			\multicolumn{1}{r}{\forall j.\ \neg ps[j].sb.\hasTagFunc(t)\ \vee\ ps[j].sb.\getOldestFunc(a) = \langle v,t \rangle)} \\
			\end{array}
		}{
			\begin{array}{ll}
			\multicolumn{2}{c}{m[a] \Leftarrow v} \\
			\forall j. & \ifFunc\ ps[j].sb.\hasTagFunc(t)\ \thenFunc\ ps[j].sb.\removeOldestFunc(a) \\ 
			& \elseFunc\ \ifFunc\ \neg ps[j].sb.\existFunc(a)\ \thenFunc\ ps[j].ib.\insertFunc(a,old) \\
			\end{array}
		}
		\end{displaymath} \reduceRuleEndSpace
		
		\textbf{\wmmSSBPropSt{} rule} (copy store from processor $i$ to $j$). \reduceRuleSpace
		\begin{displaymath}
		\frac{
			\begin{array}{c}
				\langle v, t \rangle = ps[i].sb.\getRandFunc(a) \\
				\whenFunc(t\neq \epsilon\ \wedge\ \noCycleFunc(a, t, j)) \\
			\end{array}
		}{
			ps[j].sb.\enqFunc(a,v,t);\ ps[j].ib.\removeAddrFunc(a)
		}
		\end{displaymath}
	\end{boxedminipage}
	\caption{\wmmSSB{} operational semantics} \label{fig: wmm ssb op model}
\end{figure}

In Figure \ref{fig: wmm ssb op model}, \wmmSSBStRule{} is simply introducing the store tag to the original \wmmStRule{} rule.
In \wmmSSBDeqSbRule{}, when we write a store ($\langle a, v, t\rangle$) into memory, we ensure that each copy of this store is the oldest one to address $a$ in its respective store buffers.
The old memory value is inserted into the invalidation buffer $ib$ of each processor whose $sb$ does not contain address $a$.
\wmmSSBPropSt{} copies a store ($\langle a,v,t \rangle$) from $ps[i]$ to $ps[j]$.
The check on $\noCycleFunc(a,t,j)$ guarantees that no cycle is formed in the partial coherence order after the copy.
Copying stores from $ps[i]$ to $ps[i]$ will be automatically rejected because $\noCycleFunc$ will return $\False$.
Since we enqueue a store into $ps[j].sb$, we need to remove all stale values for address $a$ from $ps[j].ib$.

The rule to execute a $\ComInst$ fence in \wmmSSB{} looks the same as that in WMM (\ie{} \wmmComRule), but has very different implications for implementations.
In \wmmSSB{}, a store cannot be moved from $sb$ to memory unless all its copies in other store buffers can be moved at the same time.
Hence the effect of a $\ComInst$ fence is no longer local; it implicitly affects all other store buffers/caches.
In literature, such fences are known as \emph{cumulative}.

\subsection{Litmus Tests for \wmmSSB}

We show by examples that \wmmSSB{} allows non-atomic multi-copy store behaviors, and that fences in \wmmSSB{} have the cumulative properties similar to those in Power and ARM memory models.

We first revisit the WWC example in Figure \ref{fig: rc no ssb}.
The behavior in the figure, which is disallowed by WMM, is now allowed by \wmmSSB.
This is because in \wmmSSB, $I_1$ could be copied into the store buffer of P2, and then $I_2$ reads its value from the store buffer.
After that, $I_3$ is written to memory, $I_4$ executes, and $I_5$ is written to memory.
Finally $I_1$ is written to memory, leading the final memory value for $a$ to be 2.
This behavior can be found in implementations in which P1 and P2 share a write-through cache.
To forbid this behavior in \wmmSSB, we can insert a $\ComInst$ fence between $I_2$ and $I_3$ on P2 to force $I_1$ to be written into memory.
The inserted $\ComInst$ fence has a cumulative global effect in ordering $I_1$ before $I_3$ (and hence $I_5$).

Figure \ref{fig: iriw wmm ssb} shows another well-known example called Independent Reads of Independent Writes (IRIW).
The non-SC behavior in the figure is allowed by \wmmSSB, while it is forbidden by the original WMM model.
This is because in \wmmSSB, $I_1$ and $I_2$ could be copied into the store buffers of P3 and P4 respectively.
Then $I_3$ and $I_6$ can read the values of $I_1$ and $I_2$ from store buffers.
After that, $I_5$ and $I_8$ simply access the monolithic memory and both get value 0.
This behavior can be found in implementations in which P1, P3 share a write-through cache, and P2, P4 share another write-through cache.

To forbid the behavior in Figure \ref{fig: iriw wmm ssb} in \wmmSSB, we can insert a $\ComInst$ fence between $I_3$ and $I_4$ on P3, and another $\ComInst$ fence between $I_6$ and $I_7$ on P4.
As we can see, a $\ComInst$ followed by a $\RecInst$ in \wmmSSB{} has the same effect as the Power $\syncInst$ fence and the ARM $\dmbInst$ fence.
Cumulation is achieved by globally advertising observed stores ($\ComInst$) and preventing later loads from reading stale values ($\RecInst$).

\begin{figure}[!htb]
	\centering
	\small
	\begin{tabular}{|l|l|l|l|}
		\hline
		Proc. P1 & Proc. P2 & Proc. P3 & Proc. P4 \\
		\hline
		$I_1: \StInst\ a\ 1$ & $I_2: \StInst\ b\ 1$ & $I_3: r_1=\LdInst\ a$ & $I_6: r_3=\LdInst\ b$ \\
		                     &                      & $I_4: \RecInst$       & $I_7: \RecInst$ \\
		                     &                      & $I_5: r_2=\LdInst\ b$ & $I_8: r_4=\LdInst\ a$ \\
		\hline
		\multicolumn{4}{|l|}{\wmmSSB{} allows: $r_1=1,\ r_2=0,\ r_3=1,\ r_4=0$} \\
		\hline
	\end{tabular}
	\caption{IRIW in \wmmSSB} \label{fig: iriw wmm ssb}
\end{figure}

\section{Conclusion} \label{sec: conclude}

Weak memory models can be tamed, that is, made more understandable without sacrificing efficiency.
One contribution to the complexity is write-through caches, and we see no fundamental advantage of such caches over write-back caches.
We do think Instantaneous Instruction Execution (\IIE) descriptions leave little room for ambiguity in the operational semantics of memory models and should be used in all definitions.
We have also presented three concrete weak memory models:
WMM, a futuristic model that is suitable when load-value prediction becomes commonplace in microarchitectures;
\wmmDep{}, a model that enforces data-dependency ordering and can be adopted immediately;
and \wmmSSB{}, an extension on WMM that allows non-atomic multi-copy store behaviors.

\section{Acknowledgement}
This work was done at CSAIL, MIT as part of the Proteus project, which is  partially funded by DARPA BRASS grant number 6933274 (2015-2019).

\bibliographystyle{ieeetr}
\bibliography{ref}

\end{document}